\documentclass[amsmath,amssymb,aip,jcp,twocolumn,10pt]{revtex4-1}

\usepackage{dcolumn}
\usepackage{graphicx}
\usepackage{graphics}
\usepackage{setspace}
\usepackage[usenames]{color}

\newcommand{\hH}{\hat{H}}
\newcommand{\hV}{\hat{V}}
\newcommand{\bv}{\bar{v}}
\newcommand{\bH}{\bar{H}}
\newcommand{\cH}{{\cal H}}

\newcommand{\hT}{\hat{T}}
\newcommand{\hR}{\hat{R}}
\newcommand{\hLe}{\hat{L}}
\newcommand{\hF}{\hat{F}}
\newcommand{\hW}{\hat{W}}
\newcommand{\bra}[1]{\langle #1|}
\newcommand{\ket}[1]{| #1\rangle}

\begin{document}

\title{Approximating electronically excited states with equation-of-motion linear coupled-cluster theory}

\author{Jason N. Byrd}
\email{byrdja@chem.ufl.edu}
\affiliation{Quantum Theory Project, University of Florida, Gainesville, FL 32611}
\author{Varun Rishi}
\affiliation{Quantum Theory Project, University of Florida, Gainesville, FL 32611}
\author{Ajith Perera}
\affiliation{Quantum Theory Project, University of Florida, Gainesville, FL 32611}
\author{Rodney J. Bartlett}
\affiliation{Quantum Theory Project, University of Florida, Gainesville, FL 32611}
\affiliation{ Max-Planck-Institut fu\"{u}r Kohlenforschung, Kaiser-Wilhelm-Platz 1, D-45470 M\"{u}̈lheim an der Ruhr, Germany}


\begin{abstract}
A new perturbative approach to canonical equation-of-motion coupled-cluster theory is presented using coupled-cluster perturbation theory.  A second-order M{\o}ller-Plesset partitioning of the Hamiltonian is used to obtain the well known equation-of-motion many-body perturbation theory (EOM-MBPT(2)) equations and two new equation-of-motion methods based on the linear coupled-cluster doubles (EOM-LCCD) and linear coupled-cluster singles and doubles (EOM-LCCSD) wavefunctions.  This is achieved by performing a short-circuiting procedure on the MBPT(2) similarity transformed Hamiltonian.  These new methods are benchmarked against very accurate theoretical and experimental spectra from 25 small organic molecules.  It is found that the proposed methods have excellent agreement with canonical EOM-CCSD state for state orderings and relative excited state energies as well as acceptable quantitative agreement for absolute excitation energies compared with the best estimate theory and experimental spectra.
\end{abstract}

\maketitle

\section{Introduction}

Optical spectroscopy is an important and ubiquitous tool in modern experimental 
studies.  Improvements over the years in the use of lasers in optical 
spectroscopy has brought the field to an impressive level of accuracy and 
precision that is difficult to match using modern theoretical methods.  Even so, 
theory has an important role to play in optical spectroscopy.  By assisting in 
excited state assignments and predicting the density of excited states 
expected in a given energy range, computational results can be a powerful tool 
for spectroscopists.

Widespread use of time dependent density functional theory\cite{dreuw2005} 
(TD-DFT) for the calculation of excited states will continue for a long time to 
come due to the attractive low computational scaling inherent to the method. 
 However, reliability and accuracy problems\cite{caricato2010} in TD-DFT keep 
the theory from replacing the more computationally expensive but highly accurate 
multireference configuration interaction\cite{szalay2012} (MRCI) or equation of 
motion coupled-cluster\cite{bartlett2007,sneskov2012} (EOM-CC) theories.  The 
most commonly 
used form of EOM-CC theory is the EOM-CCSD variant,\cite{stanton1993} which 
limits ground and excited state excitations to singles and doubles only.  This 
approximation is qualitatively consistent and approaches quantitative accuracy 
in many cases where single excitations are dominant.  
For doubly excited states, perturbative\cite{watson2013} (EOM-CCSD(T)) or iterative\cite{watts1994} 
(EOM-CCSDT-n) triple excitations are required\cite{bene1997,watts1999} to obtain quantitative agreement with experiment.

The development of approximate excited state methods based on many-body perturbation 
and coupled-cluster theory has been an active and fruitful endevour since the 
emergence of the field.\cite{paldus1978}  Many approaches make use of a 
mean-field starting reference based on configuration interaction 
singles\cite{forseman1992} and add perturbative corrections to that 
reference.\cite{headgordon1994,hirata2005,liu2013,liu2014a, liu2014b,byrd2014-b} 
 Other approaches use a many-body treatment of propagator theory such as the 
algebraic-diagrammatic construction\cite{schirmer1982,trofimov2006} (ADC) scheme 
or by using approximate coupled-cluster 
linear-response\cite{monkhorst1977,sekino1984,koch1990} theory with the CCn 
methods.\cite{christiansen1995}  Approximate methods based on equation-of-motion theory can either add a {\it post hoc} EOM-CC perturbative correction\cite{hirata2001a,hirata2001b,watson2013}for triples or
an {\it a priori} perturbative approximation.\cite{stanton1995,nooijen1995,gwaltney1996}  We will use
perturbation theory here to develop a new perturbative EOM-CC method that also 
includes infinite-order effects in the ground state wavefunction.  The goal will 
be to provide a consistent and accurate way to obtain excitation energies from 
either a linear CCD (LCCD) or CCSD (LCCSD) ground state wavefunction.  These 
methods are not intended to be a replacement method for existing fast 
(EOM-MBPT(2)) or accurate (EOM-CCSD) methods when considering just the 
calculation of excited states, but to provide a route toward a consistent excited 
state spectra from a specific ground state wavefunction.

In this work, we will use coupled-cluster perturbation theory\cite{bartlett2010} (CCPT) to derive a general EOM-CCPT framework in Sec. \ref{theory}.  The various EOM-CC approximations will then be systematically derived from EOM-CCPT by a choice in the Hamiltonian perturbation partitioning.  Numerical results of these approximations compared against very accurate theoretical and experimental spectra are presented in Sec. \ref{discussion}. 

\section{\label{theory}Theory}

\subsection{Coupled-Cluster Theory}

Central to coupled-cluster theory\cite{bartlett2007} is the similarity transformation of the electronic Hamiltonian by the exponential wave operator 
\begin{equation}\label{hbar1}
\bH=e^{-\hT}\hH e^{\hT}.
\end{equation}
The excitation cluster operators (here limited to single and double excitations) are given by 
\begin{equation}
\hT = \hT_1 + \hT_2 =
\sum_{ai} t^{a}_{i}\hat{a}^\dag\hat{i} +
\frac{1}{4} \sum_{abij}t^{ab}_{ij}\hat{a}^\dag\hat{i} \hat{b}^\dag \hat{j}
\end{equation}
and the normal ordered electronic Hamiltonian is
\begin{align}
\hH &= 
\langle \phi_0|\hat{H}|\phi_0\rangle +
\sum_{pq}f^p_q \lbrace p^\dag q\rbrace + 
\frac{1}{4}\sum_{pqrs}\bar{v}^{pq}_{rs} \lbrace p^\dag q^\dag sr\rbrace \\ \label{Nhamiltonian}
 &= \langle \phi_0|\hat{H}|\phi_0\rangle + \hF + \hW.
\end{align}
Here $f$ is the one-electron Fock matrix, 
$\bar{v}$ the antisymmetrized two-electron integrals, 
$\lbrace\cdots\rbrace$ denotes normal ordering of the enclosed operators 
and the one- ($\hF$) and two-particle ($\hW$) are defined appropriately.
Throughout we reserve the indices $i,j,k,\cdots$ and $a,b,c,\cdots$ for
occupied and virtual orbitals respectively while the indices $p,q,\cdots$ may
refer to either occupied or virtual orbitals.  Because the Hamiltonian only contains one- and two-particle operators, the Baker-Campbell-Hausdorff expansion of Eq. \ref{hbar1} naturally truncates after four commutators giving
\begin{multline}\label{hbar}
\bH=
\hH +
\left[
\hH,\hT\right]
+
\frac{1}{2!}\left[ \left[
\hH,\hT\right]
,\hT\right]
+\\
\frac{1}{3!}\left[ \left[ \left[
\hH,\hT\right]
,\hT\right]
,\hT\right]
+
\frac{1}{4!}\left[ \left[ \left[ \left[
\hH,\hT\right]
,\hT\right]
,\hT\right]
,\hT\right].
\end{multline}

Starting with an appropriate  single-reference mean-field reference, $\ket{\phi_0}$, 
the coupled-cluster Schr\"{o}dinger equation is expressed with Eq. \ref{hbar} as
\begin{equation}\label{sch1}
\bH\ket{\phi_0} = E_{\rm CC}\ket{\phi_0}.
\end{equation}
The CC correlation energy given by projecting on the left by the reference function
\begin{equation}\label{ccen}
\bra{\phi_0}\bH\ket{\phi_0}=E_{\rm CC}
\end{equation}
while the necessary cluster amplitudes are obtained by solving the equations generated by 
projecting on the left with the auxiliary space 
\begin{equation}\label{Tamp}
\bra{\phi_g}\bH\ket{\phi_0} = 0
\end{equation}
where $\bra{\phi_g}$ is the set of $g$-fold excited determinants 
\begin{equation}
\bra{\phi_g} = \bra{\phi_0}\hat{i}^{\dag}_1 \hat{a}_1 \hat{i}^{\dag}_2 \hat{a}_2\cdots \hat{i}^{\dag}_g \hat{a}_g
\end{equation}
The conventional approach to computing excited states through the equation-of-motion (EOM) method,
\cite{bartlett2007} which directly computes the $k$'th excitation energy ($\omega_k$) relative to 
the ground state coupled-cluster wavefunction, is also obtained with Eq. \ref{hbar} as
\begin{equation}\label{eomeqn}
\bra{\phi_0}\hLe(k)\left[\bH,\hR(k)\right]\ket{\phi_0}_C = 
\omega_k.
\end{equation}
Here $\omega_k$ is the $k$'th general eigenvalue, and $\hR(k)$ ($\hLe(k)$) is the $k$'th right (left)
 general eigenvector solution to the non-Hermitian matrix, $\bH$.  The $\langle\rangle_C$ denotes that 
 only fully connected contractions are included.
These eigenvector solutions are defined as the linear excitation and de-excitation operators (limited again to only singles and doubles)
\begin{multline}
\hR(k) = r_0(k) + \hR_1(k) + \hR_2(k) = \\
r_0(k) + 
\sum_{ai} r^{a}_{i}(k)\hat{a}^\dag\hat{i} +
\frac{1}{4} \sum_{abij}r^{ab}_{ij}(k)\hat{a}^\dag\hat{i} \hat{b}^\dag \hat{j}
\end{multline}
\begin{multline}
\hLe(k) = \hLe_1(k) + \hLe_2(k) =  
\sum_{ai} \ell^{i}_{a}(k)\hat{i}^\dag\hat{a} +
\frac{1}{4} \sum_{abij} \ell^{ij}_{ab}(k)\hat{i}^\dag\hat{a}\hat{j}^\dag\hat{b},
\end{multline}
with the biorthogonalization constraint
\begin{equation}
\bra{\phi_0}\hLe(k)\hR(k')\ket{\phi_0}\equiv \delta_{kk'}.
\end{equation}
Note that because $\hR(k)$ is an excitation operator, it will commute with the ground state cluster amplitudes
\begin{equation}
\left[\hT,\hR(k)\right]=0.
\end{equation}

\subsection{Coupled-Cluster Perturbation Theory Similarity Transformed Hamiltonian}

As is always possible in Hamiltonian based perturbation theory we can represent
the complete Hamiltonian in terms of a zero'th order Hamiltonian $\hH_0$ and
perturbation $\hV$ 
\begin{equation}\label{pH}
\hH = \hH_0 +\alpha \hV
\end{equation}
where $\alpha\rightarrow 1$ is an order parameter used for convenience.
Equation \ref{pH} can be directly inserted into in any coupled-cluster similarity transformation or energy functional and expanded to the desired order in $\alpha$. Examples other than the usual form given by Eq. \ref{hbar} would include 
the Hermitian\cite{paldus1972,bartlett1988} ($e^{\hT^\dag}\hH e^{\hT}$) or
unitary\cite{bartlett1989} ($e^{\hat{\tau}^\dag}\hH e^{\hat{\tau}}$) expansions, but they do not simply terminate without truncation.

To facilitate the decomposition into a reference and perturbation operator, we rearrange the Hamiltonian into particle excitation rank form 
\begin{equation}
\hH =
\langle 0|\hat{H}|0\rangle +
\hF^{[0]} +\hF^{[\pm 1]} +
\hW^{[0]} + \hW^{[\pm1]} + \hW^{[\pm2]}
\end{equation}
where the $[n]$ superscript denotes that the operator changes particle rank from right to left by $n$.
From this point the CC perturbation framework used is the same as we have employed in the past.\cite{bartlett2010,byrd2014-b,byrd2015-a}  Here the exponential wavefunction
$e^{\hT}$ is expanded to the appropriate order in terms of $n$'th order cluster
operators $\hT^{(n)}$.  The computation of these perturbative cluster operators
is done order by order using standard perturbation theory.  This is easily
illustrated by expanding Eq. \ref{hbar} using the partitioned Hamiltonian of Eq.
\ref{pH}.  The similarity transformed Hamiltonian to arbitrary order in $\alpha$ is given by
\begin{widetext}
\begin{multline}\label{hbarn}
\bH^{(n)} \equiv
\sum_{n'}
\left[
\left(\hH_0\delta_{n,n'} + \hV\delta_{n,n'+1}\right),
\hT^{(n')}\right] +  
\frac{1}{2!}
\sum_{n'm}
\left[\left[
\left(\hH_0\delta_{n,n'+m} + \hV\delta_{n,n'+m+1}\right),
\hT^{(m)}\right],\hT^{(n')}\right] 
+ \\
\frac{1}{3!}
\sum_{n'mm'}
\left[\left[\left[
\left(\hH_0\delta_{n,n'+m+m'} + \hV\delta_{n,n'+m+m'+1}\right),
\hT^{(m)}\right],\hT^{(m')}\right],\hT^{(n')}\right]
+ \\
\frac{1}{4!}
\sum_{n'mm'm''}
\left[\left[\left[\left[
\left(\hH_0\delta_{n,n'+m+m'+m''} + \hV\delta_{n,n'+m+m'+m''+1}\right),
\hT^{(m)}\right],\hT^{(m')}\right],\hT^{(m'')}\right],\hT^{(n')}\right].
\end{multline}
\end{widetext}
Throughout this work we denote the exclusive order in $\alpha$ of an operator 
with the superscript $(n)$ and we use $\lbrace n\rbrace$ to denote an expansion 
inclusive of all orders in $\alpha$ up to $n$.  With this notation, the general 
$n$'th 
order similarity transformed Hamiltonian can be 
expressed as a compact series
\begin{equation}\label{hbarninclusive}
\bH^{\lbrace n\rbrace} = \hH_0 + \bH^{(1)} + \cdots + \bH^{(n)}.
\end{equation}

The $n$'th order ground state CCPT cluster amplitudes are completely defined from  insertion of the exclusive similarity transformed Hamiltonian, Eq. \ref{hbarn}, into \ref{Tamp}  
\begin{equation}\label{ccptTamp}
\bra{\phi_g}\bH^{(n)}\ket{\phi_0} = 0.
\end{equation}
Using these $n$'th order ground state amplitudes in Eq. \ref{ccen} will give the $n+1$ order correlation energy beginning with $n=2$
\begin{equation}\label{ccpten}
\bra{\phi_0}\bH^{(n+1)}\ket{\phi_0}=E^{(n+1)}_{\rm CC}.
\end{equation}
From this the exact coupled-cluster correlation energy, relative to the mean-field energy
\begin{equation}\label{cccorren}
\Delta E_{\rm CC} =
E_{\rm CC} - \langle 0|\hat{H}|0\rangle,
\end{equation}
is
\begin{equation}
\Delta E_{\rm CC} \simeq 
\Delta E^{(2)}_{\rm CC} + 
\Delta E^{(3)}_{\rm CC} + 
\dots +
\Delta E^{(n)}_{\rm CC}
\end{equation}
or equivalently 
\begin{align}\label{ccptentotal}
\Delta E_{\rm CC} &\simeq  
\bra{\phi_0}\bH^{(2)}\ket{\phi_0} + 
\dots +
\bra{\phi_0}\bH^{(n)}\ket{\phi_0}\\
&\equiv
\bra{\phi_0}\bH^{\lbrace n\rbrace}\ket{\phi_0}.
\end{align}
To obtain excitation energies relative to the $n$'th order energy without perturbatively expanding the $\hLe$ and $\hR$ operators, we use the $n$'th order {\it inclusive} similarity transformed Hamiltonian (Eq. \ref{hbarninclusive}) in Eq. \ref{eomeqn} rather than the {\it exclusive} Eq. \ref{hbarn}, giving
\begin{equation}\label{ccpteomeqn}
\bra{\phi_0}\hLe(k)\left[\bH^{\lbrace n\rbrace},\hR(k)\right]\ket{\phi_0} = 
\omega_k.
\end{equation}
There are many ways to partition the Hamiltonain.  We have used the particle rank conserving partitioning of the Hamiltonian,
\begin{align}
\hH_0 &= \langle 0|\hat{H}|0\rangle + \hF^{[0]} + \hW^{[0]} \\
\hV   &= \hF^{[\pm 1]} + \hW^{[\pm1]} + \hW^{[\pm2]},
\end{align}
in previous work\cite{bartlett2010,byrd2014-b,byrd2015-a} to great success.  Alternatively we can also choose the traditional M{\o}ller-Plesset (MP) partitioning which we will discuss next.

\subsection{Second-Order M{\o}ller-Plesset Partitioning}

\begin{figure}
\center\includegraphics[width=0.95\columnwidth]{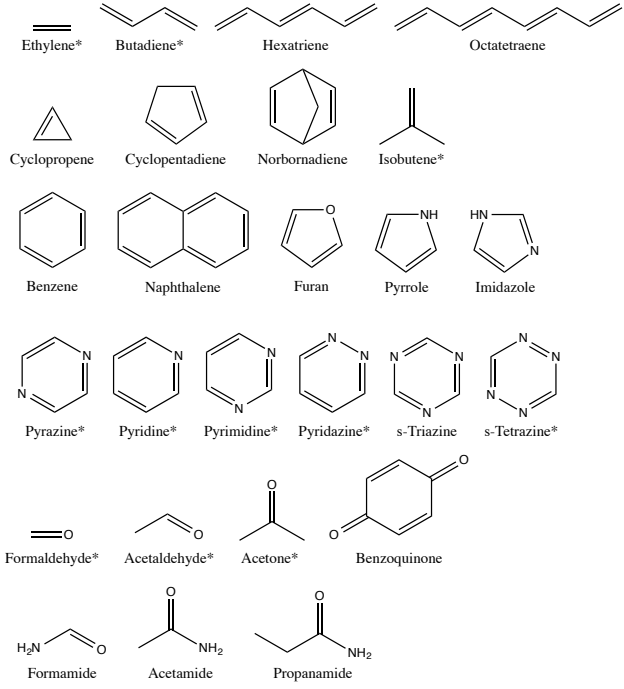}
\caption{\label{molecules}Member molecules of the M\"{u}lheim test set.  Names denote with an ${}^*$ are also members of the Yale set, with Acetaldehyde only a member of the Yale set.}
\end{figure}

The standard approach in electronic structure theory is to use generalized many-body perturbation theory (GMBPT),\cite{shavitt2009} based on the M{\o}ller-Plesset (MP) partitioning of the Hamiltonian, given by
\begin{align}
\label{h0mp}
\hH_0 &= \hat{F}^{[0]} \\
\label{vmp}
\hV   &= \hat{F}^{[\pm 1]} + \hat{W}^{[0]} + \hat{W}^{[\pm 1]} + \hat{W}^{[\pm 2]}.
\end{align}
Using Eqs. \ref{h0mp} and \ref{vmp} in Eq. \ref{hbarn}, the first two orders of $\bH^{(n)}$ are respectively
\begin{equation}
\label{mphbar1}
\bH^{(1)} =
\left[\hH_0,\hT^{(1)}\right] + \hV
\end{equation}
and
\begin{equation}
\label{mphbar2}
\bH^{(2)} =
\left[\hH_0,\hT^{(2)}\right] + 
\left[\hV,\hT^{(1)}\right].
\end{equation}
Inserting Eq. \ref{mphbar1} into Eq. \ref{ccptTamp} gives the quintessential GMBPT(1) wavefunction
\begin{align}
\label{mp2wfSingles}
\bra{\phi_1} \hat{F}^{[0]} \hT^{(1)}_1 + F^{[+1]} \ket{\phi_0}_C &= 0 \\
\label{mp2wfDoubles}
\bra{\phi_2} \hat{F}^{[0]} \hT^{(1)}_2 + W^{[+2]} \ket{\phi_0}_C &= 0 
\end{align}
from which 
\begin{equation}
E^{(2)} = \bra{\phi_0}\hF^{[-1]}\hT^{(1)}_1 + \hW^{[-2]}\hT^{(1)}_2\ket{\phi_0}_C.
\end{equation}
The GMBPT(2) wavefunction is
\begin{equation}
\label{mp3wfSingles}
\bra{\phi_1} \hat{F}^{[0]} \hT^{(2)}_1 + \hat{W}^{[0]} \hT^{(1)}_1 + (\hF^{[-1]} + \hW^{[-1]}) \hT^{(1)}_2 \ket{\phi_0}_C = 0
\end{equation}
\begin{equation}
\label{mp3wfDoubles}
\bra{\phi_2} \hat{F}^{[0]} \hT^{(2)}_2 + \hat{W}^{[0]} \hT^{(1)}_2 + W^{[+1]} \hT^{(1)}_1 \ket{\phi_0}_C = 0.
\end{equation}
The triples contribution does not contribute to the first- or second-order wavefunction.  
The inclusive second-order similarity transformed Hamiltonian,
\begin{equation}\label{mphbarinclusive}
\bH^{\lbrace 2\rbrace} =
\hat{F}^{[0]} + 
\hV + 
\left[\hat{F}^{[0]},\hT^{(1)}\right] + 
\left[\hat{F}^{[0]},\hT^{(2)}\right] + 
\left[\hV,\hT^{(1)}\right],
\end{equation}
is the GMBPT generalization of previous\cite{stanton1995,nooijen1995,gwaltney1996} MBPT approximations to $\bH$.  The spin-orbital  equations for the $\bH^{\lbrace 2\rbrace}$ matrix elements\cite{stanton1995,nooijen1995} using Einstein notation are
\begin{align}
\label{MPij}
\cH^{i}_{j} =& 
    f^i_j - 
    f^{i}_{e}t^{e(1)}_{j} -
    \bv^{im}_{je} t^{e(1)}_{m} +
    \frac{1}{2} \bv^{im}_{ef} t^{ef(1)}_{jm} 
,\\
\label{MPab}
\cH^{a}_{b} =&
    f^{a}_{b} +
    f^{m}_{b} t^{a(1)}_{a} + 
    \bv^{am}_{be} t^{e(1)}_{m} + 
    \frac{1}{2} \bv^{mn}_{be} t^{ae(1)}_{mn}
,\\
\label{MPia}
\cH^{i}_{a} =&
    f^i_a + 
    \bv^{im}_{ae} t^{e(1)}_{m}
,\\
\label{MPai}
\cH^{a}_{i} =& 0
,
\end{align}
\begin{align}
\label{MPabcd}
\cH^{ab}_{cd} =&
    \bv^{ab}_{cd} + 
    \bv^{am}_{cd} t^{b(1)}_{m} +
    \frac{1}{2} \bv^{mn}_{cd} t^{ab(1)}_{mn}
,\\
\label{MPijkl}
\cH^{ij}_{k\ell} =&
    \bv^{ij}_{k\ell} + 
    \bv^{ij}_{ke}t^{e(1)}_{\ell} +
    \frac{1}{2} \bv^{ij}_{ef} t^{ef(1)}_{k\ell}
,\\
\label{mPiabj}
\cH^{ia}_{bj} =&
    \bv^{ia}_{bj} -
    \bv^{mi}_{bj} t^{a(1)}_{m} +
    \bv^{ia}_{be} t^{e(1)}_{j} +
    P(ij)P(ab) \bv^{mi}_{eb} t^{ae(1)}_{mj}
,\\
\label{MPijka}
\cH^{ij}_{ka} =&
    \bv^{ij}_{ka} - 
    \bv^{ij}_{ea} t^{e(1)}_{k}
,\\
\label{MPaibc}
\cH^{ai}_{bc} =&
    \bv^{ai}_{bc} + 
    \bv^{mi}_{bc} t^{a(1)}_{m}
,\\
\nonumber
\cH^{ia}_{jk} =&
    \bv^{ia}_{jk} + 
    \bv^{im}_{jk} t^{a(1)}_{m} - 
    \bv^{ia}_{je} t^{e(1)}_{k} + 
    f^{i}_{e} t^{ea(1)}_{jk} + \\ \label{MPiajk} &
    \frac{1}{2} \bv^{ia}_{ef} t^{ef(1)}_{jk} - 
    P(jk) \bv^{im}_{je} t^{ea(1)}_{mk}
,\\
\nonumber
\cH^{ab}_{ci} =& 
    \bv^{ab}_{ci} -
    \bv^{ab}_{ce} t^{e(1)}_{i} +
    \bv^{am}_{ci} t^{b(1)}_{m} - \\ \label{MPabci} &
    \frac{1}{2} \bv^{mn}_{ci} t^{ab(1)}_{mn} +
    P(ab) \bv^{am}_{ce} t^{eb(1)}_{mi}
,\\
\label{MPijab}
\cH^{ij}_{ab} =& \bv^{ij}_{ab}
,\\
\label{MPabij}
\cH^{ab}_{ij} =& 0
,
\end{align}
\begin{align}
\label{MPajbcdi}
\cH^{ajb}_{cdi} =&
    \bv^{mj}_{cd}  t^{ab(1)}_{mi}
,\\
\label{MPijakbl}
\cH^{ija}_{kb\ell} =&
    -\bv^{ij}_{be} t^{ea(1)}_{k\ell}
,\\
\label{MPiabcjk}
\cH^{iab}_{cjk} =&
    P(ab) \bv^{ia}_{ce} t^{eb(1)}_{ij}-
    P(ij) \bv^{im}_{cj} t^{ab(1)}_{mk},
\end{align}
where $P(pq)$ is the anti-permutation operator
\begin{equation}
P(pq) g(pq\dots rs) = g(pq\dots rs) - g(pq\dots sr).
\end{equation}
The difference between GMBPT and MP theory occurs when $f^a_i$, $f^i_j$ and $f^a_b\ne 0$ as they would be for canonical HF orbitals.  The first  introduces a first-order singles wavefunction (Eq. \ref{mp2wfSingles}), while the second and third terms have to be summed to all order as in coupled-cluster theory or better transformed away by a semi-canonical occupied-occupied and virtual-virtual rotation.  
The first-order $t$ amplitudes used in Eqs. \ref{MPij}-\ref{MPiabcjk} are obtained from Eq. \ref{mp2wfSingles} and Eq. \ref{mp2wfDoubles}.  Second-order amplitude contributions are limited in this case to $\cH^{a}_{i}$ and $\cH^{ab}_{ij}$ because in the GMBPT partitioning $\hat{F}^{[0]} T^{(2)}$ can only give particle excitation matrix elements.  Because these matrix elements are by definition zero (Eq. \ref{Tamp}) there are no actual contributions from the second-order amplitudes to the excited state spectrum.

\subsection{Inclusive Second-Order M{\o}ller-Plesset Partitioning}

\begin{figure*}
\center\includegraphics[width=1.95\columnwidth]{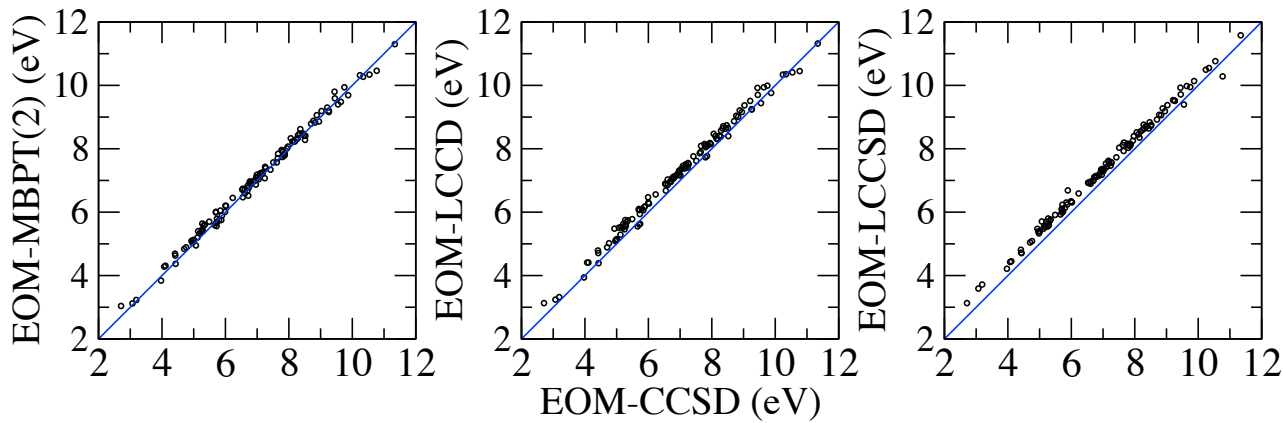}
\caption{\label{vsccsd}Comparison against the M\"{u}lheim data set EOM-CCSD values.}
\end{figure*}

The EOM-MBPT(2) approach theoretically suffers from the fact that the starting wavefunction is strictly a single shot perturbation calculation, while the diagonalization of $\bH$ introduces infinite-order contributions to the excitation wavefunction.  This can be overcome, while using the GMBPT partitioning, by short-circuiting the order-by-order amplitude computation to instead solve
\begin{equation}
\bra{\phi_g}\bH^{\lbrace 2\rbrace}\ket{\phi_0} = 0.
\end{equation}
Here we combine the first- and second-order amplitude equations to obtain
\begin{multline}
\label{infmp2wfSingles}
\bra{\phi_1} 
\hat{F}^{[0]} \hT^{(1\infty)}_1 + \hat{W}^{[0]} \hT^{(1\infty)}_1 + 
\hF^{[-1]}\hT^{(1\infty)}_2 + \\ \hW^{[-1]} \hT^{(1\infty)}_2 + F^{[+1]} 
\ket{\phi_0}_C = 0
\end{multline}
\begin{multline}
\label{infmp2wfDoubles}
\bra{\phi_2} 
\hat{F}^{[0]} \hT^{(1\infty)}_2 + \hat{W}^{[0]} \hT^{(1\infty)}_2 + \\
\hW^{[+1]} \hT^{(1\infty)}_1 + \hW^{[+2]}
\ket{\phi_0}_C = 0.
\end{multline}
Equations \ref{infmp2wfSingles} and \ref{infmp2wfDoubles} are incidentally the linearized CCSD amplitude equations (LCCSD).  If the singles contributions in Eq. \ref{infmp2wfSingles} are removed, the resulting reference wavefunction is then simply linearized CCD (LCCD).\cite{taube2009}  This is a convenient approximation as it requires no new terms in the similarity transformed Hamiltonian, while introducing infinite-order character to the reference wavefunction.

\section{\label{esc}Electronic Structure Calculations}

\begin{figure*}
\center\includegraphics[width=1.95\columnwidth]{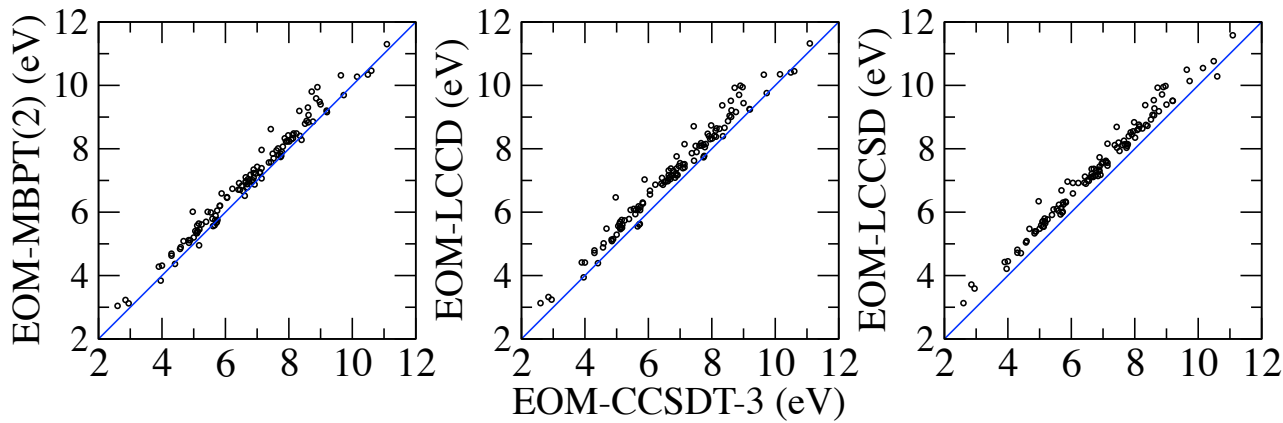}
\caption{\label{vsccsdt3}Comparison against the M\"{u}lheim data set EOM-CCSDT-3 values.}
\end{figure*}

\begin{table}
\begin{ruledtabular}
\begin{tabular}{lccccc}
  State & MBPT(2) & LCCD & LCCSD & CCSD & Exp.\\
\hline
\multicolumn{4}{l}{ethylene}\\
 $ B_{3u }$& 7.15 & 7.43 & 7.45 & 7.28 & 7.11\\
 $ B_{1u }$& 7.76 & 8.13 & 8.18 & 7.97 & 7.65\\
 $ B_{1g }$& 7.79 & 8.08 & 8.10 & 7.97 & 7.80\\
 $ B_{2g }$& 7.84 & 8.12 & 8.14 & 7.93 & 7.90\\
 $ A_{g }$& 8.19 & 8.46 & 8.48 & 8.31 & 8.28\\
 $ B_{3u }$& 8.64 & 8.92 & 8.94 & 8.77 & 8.62\\
 $ B_{3u }$& 8.93 & 9.20 & 9.22 & 9.05 & 8.90\\
 $ B_{3u }$& 9.05 & 9.33 & 9.35 & 9.17 & 9.08\\
 $ B_{1g }$& 9.19 & 9.46 & 9.49 & 9.34 & 9.20\\
 $ B_{1u }$& 9.18 & 9.47 & 9.50 & 9.32 & 9.33\\
 $ B_{3u }$& 9.84 & 10.11 & 10.13 & 9.96 & 9.51\\
\\
\multicolumn{4}{l}{isobutene} \\
 $ B_{1 }$& 6.29 & 6.54 & 6.56 & 6.38 & 6.17\\
 $ A_{1 }$& 6.81 & 7.06 & 7.10 & 6.91 & 6.70\\
\\
\multicolumn{4}{l}{trans-1,3-butadiene}\\
 $ B_{u }$& 6.12 & 6.54 & 6.63 & 6.29 & 5.91\\
 $ B_{g }$& 6.17 & 6.47 & 6.49 & 6.24 & 6.22\\
 $ A_{u }$& 6.47 & 6.79 & 6.80 & 6.55 & 6.66\\
 $ B_{u }$& 7.06 & 7.39 & 7.43 & 7.16 & 7.07\\
 $ B_{g }$& 7.28 & 7.59 & 7.60 & 7.36 & 7.36\\
 $ A_{g }$& 7.54 & 7.85 & 7.85 & 7.61 & 7.62\\
 $ B_{u }$& 8.08 & 8.39 & 8.40 & 8.15 & 8.00\\
 \hline
MAD & 0.09 & 0.31 & 0.34 & 0.14 \\
RMS & 0.12 & 0.34 &	0.37 & 0.19\\
MD & 0.33 & 0.63 &	0.72 & 0.45 
\end{tabular}
\end{ruledtabular}
\caption{\label{yalealkene}Excitation energies for a few small alkene molecules computed with the d-aug-cc-pVDZ basis using MP2/6-311+G** geometries.  Units are in electron volts (eV), experimental references can be found in Ref. \onlinecite{caricato2010}.}
\end{table}

To assess both the quality and general behavior of these EOM approximations when 
computing vertical excitation energies, we employ two established sets of gas 
phase molecules (see Figure \ref{molecules}): the set of 24 organic molecules 
from Schreiber {\it et al.}\cite{schreiber2008,sauer2009,silva-junior2010} 
(referred to hereafter as the  M\"{u}lheim set) and the set of 11 organic 
molecules from Caricato {\it et al.}\cite{caricato2010,caricato2011} (referred 
to here as the Yale set).  This M\"{u}lheim set contains 121 reference single 
excitation energies\cite{schreiber2008} which use molecular geometries obtained 
with an MP2/6-31G* optimization in which the first row $1s$ atomic orbitals are 
dropped.  The extent and balanced nature of the  M\"{u}lheim set readily lends 
itself to obtaining statistics between different theoretical methods.  To 
compare the approximate EOM-CC methods proposed here, we take our best 
theoretical reference to be EOM-CCSDT-3,\footnote{
Partial inclusion of 
triples\cite{watts1996} which has all possible $\hT_1$ and $\hT_2$ to $\hT_3$ 
with $O(n^7)$ scaling, but neglects the $\hT_3\rightarrow\hT_3$ contributions 
that would introduce an $O(n^8)$ step.
}
 which is known to give both accurate 
results and is a systematic EOM-CC type theory which will provide a balanced 
comparison.  The Yale set choice in molecules is a subset of the M\"{u}lheim set 
with the simple addition of acetaldehyde, but with a different selection in 
excitation energies (69 in all) whose reference values are taken from gas phase 
experiments (the experiment literature references can be found in Ref. 
\onlinecite{caricato2010}).  This comparison to experiment provides a different 
and valuable perspective with which to augment theory vs. theory examinations.  
Geometries for the Yale set are obtained from an MP2/6-311+G** optimization.

Calculations using the M\"{u}lheim set typically\cite{schreiber2008} use the 
TZVP basis set ($3s1p$ on hydrogen $5s3p1d$ on first row atoms) from Sch\"{a}fer 
{\it et al.}\cite{schafer1992}  This is a light weight basis set that allows the 
use of many different kinds of computational methods with only a modest cost in 
overall accuracy.\cite{silva-junior2010}  While perfectly acceptable when 
comparing different methods to each other, this basis is not accurate enough to 
compare with experimental results.  This is especially true when calculating 
Rydberg states, where the inclusion of many diffuse functions is 
key.\cite{gwaltney1995,wiberg2002}  When computing the excitation energies for the Yale set, 
we use the doubly diffuse function augmented Dunning 
basis,\cite{dunning1989,kendall1992} d-aug-cc-pVDZ ($4s3p$ on hydrogen and 
$5s4p3d$ on first row atoms).  This basis is comparable to the 6-311(3+,3+)G** 
basis used by Caricato {\it et al.} in their 
calculations\cite{caricato2010,caricato2011,goings2014} with almost no gain in 
including a third set of diffuse functions.\footnote{Where the d-aug-cc-pVDZ 
basis is clearly not as accurate as the 6-311(3+,3+)G** basis is in the case of 
the highest considered excited state of ethylene and formaldehyde.}

All of the reported electronic structure results were obtained using the serial 
ACESII\cite{acesii} and parallel Aces4\cite{aces4-electronic} {\it ab initio} 
quantum chemistry packages.  Calculations were performed on the University of 
Florida HiPerGator high performance cluster.  Throughout this paper we use RMS 
to mean "root mean square," MAD to mean "mean average absolute deviation," and 
MD to mean "absolute maximum deviation."

\section{\label{discussion}Numerical Results and Discussion}

\begin{table}
\begin{ruledtabular}
\begin{tabular}{lccccc}
  State & MBPT(2) & LCCD & LCCSD & CCSD & Exp.\\
\hline
\multicolumn{4}{l}{acetaldehyde}\\
 $ A''$ & 4.19 & 4.28 & 4.57 & 4.32 & 4.28\\
 $ A'$ & 6.64 & 6.68 & 6.94 & 6.78 & 6.82\\
 $ A'$ & 7.32 & 7.59 & 7.61 & 7.67 & 7.46\\
 $ A'$ & 7.56 & 7.37 & 7.82 & 7.46 & 7.75\\
 $ A'$ & 8.24 & 8.27 & 8.51 & 8.36 & 8.43\\
 $ A'$ & 8.27 & 8.31 & 8.54 & 8.39 & 8.69\\
\\
\multicolumn{4}{l}{acetone}\\
 $ A_{2}$ & 4.38 & 4.44 & 4.75 & 4.48 & 4.43\\
 $ B_{2}$ & 6.25 & 6.29 & 6.54 & 6.37 & 6.36\\
 $ A_{2}$ & 7.17 & 7.19 & 7.44 & 7.27 & 7.36\\
 $ A_{1}$ & 7.26 & 7.29 & 7.55 & 7.37 & 7.41\\
 $ B_{2}$ & 7.25 & 7.27 & 7.51 & 7.35 & 7.49\\
 $ A_{1}$ & 7.88 & 7.90 & 8.15 & 7.98 & 7.80\\
 $ B_{2}$ & 7.65 & 7.68 & 7.93 & 7.76 & 8.09\\
 $ B_{1}$ & 7.95 & 7.98 & 8.22 & 8.06 & 8.17\\
\\
\multicolumn{4}{l}{formaldehyde}\\
 $ A_{2}$ & 3.82 & 3.96 & 4.23 & 3.99 & 4.00\\
 $ B_{2}$ & 6.87 & 6.93 & 7.17 & 7.03 & 7.08\\
 $ B_{2}$ & 7.70 & 7.75 & 7.96 & 7.83 & 7.97\\
 $ A_{1}$ & 7.83 & 7.87 & 8.10 & 7.97 & 8.14\\
 $ A_{2}$ & 8.06 & 8.09 & 8.32 & 8.19 & 8.37\\
 $ B_{2}$ & 8.77 & 8.82 & 9.05 & 8.91 & 8.88\\
 $ B_{1}$ & 9.10 & 9.14 & 9.05 & 9.23 & 9.00\\
 $ A_{2}$ & 9.20 & 9.24 & 9.45 & 9.32 & 9.22\\
 $ B_{2}$ & 9.03 & 9.07 & 9.29 & 9.16 & 9.26\\
 $ A_{1}$ & 9.07 & 9.12 & 9.35 & 9.20 & 9.58\\
 $ B_{2}$ & 9.07 & 9.10 & 9.32 & 9.19 & 9.63\\
 \hline
MAD &   0.22&	0.18&	0.14&	0.13\\
RMS &   0.26&	0.23&	0.17&	0.18\\
MD &   0.56&	0.53&	0.35&	0.44
\end{tabular}
\end{ruledtabular}
\caption{\label{yalecarbonyl}Excitation energies for a few small carbonyl molecules computed with the d-aug-cc-pVDZ basis using MP2/6-311+G** geometries.  Units are in electron volts (eV), experimental references can be found in Ref. \onlinecite{caricato2010}.}
\end{table}

\subsection{EOM-MBPT(2)}

Developed two decades ago for canonical HF\cite{stanton1995} EOM-MBPT(2) is a well established 
perturbation method which has been largely ignored by the community in favor of other approximations such as 
CC2\cite{christiansen1995} or ADC(2).\cite{schirmer1982} 
 The EOM-MBPT(2) method remains relevant because of the non-iterative $O(n^5)$ 
computational cost of the ground-state wavefunction and the general accuracy of MBPT(2).  It is also reasonably accurate 
for equilibrium vertical excitation energies.  This accuracy, within the 
M\"{u}lheim test set, is illustrated in the correlation figures \ref{vsccsd}(a), 
\ref{vsccsdt3}(a) and Table \ref{mullheimstats}.  Compared to the standard 
EOM-CCSD method, EOM-MBPT(2) has an average deviation of {$0.13$} eV 
while comparison with EOM-CCSDT-3 gives an average deviation of 
{$0.31$} eV respectively, consistent with the known $\sim 0.2$ eV triples correction to EOM-CCDSD.  


The EOM-MBPT(2) approximation and its L\"{o}wdin partitioned 
variant\cite{lowdin1963,gwaltney1996} was recently benchmarked (using the 
6-311(3+,3+)G** basis set) against EOM-CCSD and experiment using the Yale test 
set by Goings {\it et al.}\cite{goings2014}  
Our computed EOM-MBPT(2)/d-aug-cc-pVDZ values, presented in Tables 
\ref{yalealkene}, \ref{yalecarbonyl} and \ref{yaleazabenzene}, yield nearly 
identical results with minor differences due to the alternative basis set 
choice.  The EOM-MBPT(2) method is completely satisfactory when compared to 
experimental Rydberg states (such as with the alkenes and carbonyl's), but is 
much less reliable for the $n\rightarrow\pi^*$ and $\pi\rightarrow\pi^*$ 
heterocycle valence states.  The comparative difficulty in describing valence states is a 
limitation of EOM theory in general which is only satisfactorily overcome in CC 
theory by using methods that include triples or other methods such as similarity transformed equation-of-motion 
coupled-cluster theory\cite{nooijen1997a,nooijen1997b,nooijen1997c,sous2014} 
(STEOM-CC).

\begin{table}
\begin{ruledtabular}
\begin{tabular}{lccccc}
  State & MBPT(2) & LCCD & LCCSD & CCSD & Exp.\\
\hline
\multicolumn{4}{l}{pyrazine}\\
 $B_{3u}$ & 4.52 & 4.63 & 4.63 & 4.33 & 3.83\\
 $B_{2u}$ & 5.36 & 5.51 & 5.50 & 5.10 & 4.81\\
 $B_{2g}$ & 6.17 & 6.26 & 6.29 & 6.01 & 5.46\\
 $B_{1g}$ & 7.14 & 7.35 & 7.34 & 7.07 & 6.10\\
 $B_{1u}$ & 7.04 & 7.28 & 7.33 & 6.95 & 6.51\\
\\
\multicolumn{4}{l}{pyridazine}\\
 $B_{1}$ & 4.03 & 4.32 & 4.36 & 4.03 & 3.60\\
 $A_{1}$ & 5.33 & 5.33 & 5.75 & 5.33 & 5.00\\
 $A_{2}$ & 4.63 & 4.63 & 4.96 & 4.63 & 5.30\\
 $B_{1}$ & 6.62 & 6.62 & 6.93 & 6.62 & 6.00\\
 $B_{2}$ & 6.26 & 6.26 & 6.43 & 6.26 & 6.50\\
\\
\multicolumn{4}{l}{pyridine}\\
 $B_{1}$ & 5.28 & 5.38 & 5.46 & 5.17 & 4.59\\
 $B_{2}$ & 5.41 & 5.57 & 5.61 & 5.23 & 4.99\\
 $A_{2}$ & 5.64 & 5.80 & 5.92 & 5.60 & 5.43\\
 $A_{1}$ & 6.76 & 6.86 & 6.92 & 6.69 & 6.38\\
\\
\multicolumn{4}{l}{pyrimidine}\\
 $B_{1}$ & 4.74 & 4.81 & 4.97 & 4.63 & 3.85\\
 $A_{2}$ & 5.07 & 5.18 & 5.38 & 5.03 & 4.62\\
 $B_{2}$ & 5.66 & 5.77 & 5.89 & 5.47 & 5.12\\
 $A_{2}$ & 6.26 & 6.35 & 6.51 & 6.18 & 5.52\\
 $B_{1}$ & 6.53 & 6.66 & 6.83 & 6.50 & 5.90\\
 $A_{1}$ & 7.02 & 7.17 & 7.37 & 6.96 & 6.70\\
\\
\multicolumn{4}{l}{s-tetrazine}\\
 $B_{3u}$ & 2.97 & 3.08 & 3.07 & 2.67 & 2.25\\
 $A_{u}$ & 4.16 & 4.30 & 4.32 & 3.98 & 3.40\\
 $A_{u}$ & 6.00 & 6.13 & 6.11 & 5.72 & 5.00\\
 $B_{3u}$ & 7.11 & 7.25 & 7.27 & 6.95 & 6.34\\
 \hline
MAD & 0.60 & 0.72 & 0.78	&	0.49\\
RMS & 0.64 & 0.76 & 0.82	&	0.53\\
MD & 1.04 & 1.25 & 1.24	    &	0.97
\end{tabular}
\end{ruledtabular}
\caption{\label{yaleazabenzene}Excitation energies for a number of single ring azabenzenes computed with the d-aug-cc-pVDZ basis using MP2/6-311+G** geometries.  Units are in electron volts (eV), experimental references can be found in Ref. \onlinecite{caricato2010}.}
\end{table}

\subsection{Linear EOM-CC}

By employing Eqs. \ref{infmp2wfSingles} and \ref{infmp2wfDoubles} to compute the
amplitudes used in the second-order similarity transformed Hamiltonian
$\bH^{\lbrace 2\rbrace}$ (Eq. \ref{mphbarinclusive}), we now have an
equation-of-motion theory completely consistent with with the LCCD (LCCSD) wavefunction
that only includes terms in the similarity transformed Hamiltonian up to $\alpha^2$ while ensuring
\begin{equation}
\bra{\phi_g}\bar{H}\ket{\phi_0}=0.
\end{equation}  
These linear EOM methods have the same iterative $O(n^6)$
computational scaling as canonical EOM-CCSD.  However, with the removal of the
quadratic amplitude intermediates, the computational overhead and I/O demands are
significantly reduced.  Benchmark data from the M\"{u}lheim set shows a systematic
overestimate of excitation energies for both EOM-LCCD and EOM-LCCSD, with
average deviations relative to EOM-CCSD of { $0.25$ and $0.35$} eV
respectively (see the correlation figures \ref{vsccsd}, \ref{vsccsdt3} and Table
\ref{mullheimstats}).  The comparison of the linear EOM methods against EOM-CCSDT-3
is less satisfactory with a MAD of { $0.45$ and $0.56$} eV for
EOM-LCCD and EOM-LCCSD respectively. 

However, these methods are not without merit should the ground state wavefunction need to
be computed with LCCD or LCCSD.  This is an important concern for weakly interacting systems 
or transition states where an MBPT(2) wavefunction is a poor approximation.
With an acceptable standard deviation of
{ $0.11$ and $0.10$} eV compared to EOM-CCSD, the excitation energies
and relative state orderings obtained with the linear EOM methods are reliably
consistent compared to canonical EOM-CCSD.  This is further illustrated in
Tables \ref{yalealkene}, \ref{yalecarbonyl} and \ref{yaleazabenzene} where the
systematic and consistent overestimation of the excitation energies is balanced
by a nearly exact agreement in state ordering {\it and} relative state energy
differences (a MAD of $0.04$, $0.04$ and $0.05$ eV for EOM-MBPT(2),
LCCD and LCCSD respectively).  For approximate methods, getting these relative
properties correct is just as useful as quantitatively accurate excitation
energies.   

\begin{table}
\begin{ruledtabular}
\begin{tabular}{lccc}
 & MBPT(2) & LCCD & LCCSD \\
\hline
\multicolumn{4}{l}{relative to EOM-CCSD}\\
MAD  & 0.13 & 0.25 & 0.35\\
RMS  & 0.15 & 0.28 & 0.36\\
MD   & 0.37 & 0.55 & 0.80\\
\multicolumn{4}{l}{relative to EOM-CCSDT-3}\\
MAD  & 0.31 & 0.45 & 0.56\\
RMS  & 0.38 & 0.53 & 0.60\\
MD   & 1.19 & 1.49 & 1.37\\
\end{tabular}
\end{ruledtabular}
\caption{\label{mullheimstats}Excitation energy comparison statistics for the M\"{u}lheim test set.  Units are in electron volts (eV), EOM-CCSD and EOM-CCSDT-3 reference values were taken from Sous {\it et al.}\cite{sous2014}}3
\end{table}

\section{\label{conclusions}Conclusions}

We have expanded the coupled-cluster similarity transformed Hamiltonian using
general coupled-cluster perturbation theory to obtain the arbitrary order coupled-cluster perturbation theory
effective Hamiltonian given by Eq. \ref{hbarn}.  The inclusive (the sum of all orders $0$ to $n$) form of
the Hamiltonian is inserted into the standard electronic-excitation
equation-of-motion theory to give the completely general EOM-CCPT (Eq.
\ref{ccpteomeqn}).  The result is a way of using the standard equation-of-motion
theory to directly compute excitation energies that are consistent with a given
CCPT ground state wavefunction. 
 
By choosing the generalized many-body perturbation theory partitioning of the Hamiltonian, we re-derive
the well known EOM-MBPT(2)\cite{gwaltney1996} (EOM-CCSD(2)\cite{stanton1995})
equations.  An alternative source of amplitudes is obtained by short-circuiting
the inclusive $\bra{\phi_g}\bH^{\lbrace 2 \rbrace}\ket{\phi_0}$ set of
equations.  The resulting infinite-order amplitudes correspond to the linear
CCSD (and CCD in the case of single excitations being neglected) expansion, and
their use in the inclusive similarity transformed Hamiltonian in place of the
standard MBPT(2) amplitudes provides an equation-of-motion method perturbatively consistent with
a LCCD and LCCSD reference wavefunction.

These approximate equation-of-motion methods are benchmarked by employing the
M\"{u}lheim\cite{schreiber2008,sauer2009,silva-junior2010} and
Yale\cite{caricato2010,caricato2011} small organic molecule data sets.  We use
the diverse M\"{u}lheim test set to benchmark our new methods against the
canonical EOM-CCSD and EOM-CCSDT-3 methods, while the Yale test set is employed in
benchmarking against experimental spectra.  Our methods are found to
consistently over estimate excitation energies, relative to EOM-CCSD and
EOM-CCSDT-3 by $\sim 0.25$ and $\sim 0.5$ eV respectively. The precise
statistics and correlation plots can be found in Table \ref{mullheimstats} and
Figures \ref{vsccsd} and \ref{vsccsdt3}.  This systematic overestimation compared 
to the complete EOM-CCSD suggests that the similarity transformed Hamiltonian (and 
thus the excited state spectrum) based on the linear CC wavefunction is over-approximated 
compared to the ground-state description leading to an increased separation.
The very good $\sim 0.1$ eV standard
deviation compared with EOM-CCSD suggests that the predicted relative spectra
will be much more accurate than the absolute excitation energies.  This is
supported by studying the benchmark values from the Yale set given in Tables 
\ref{yalealkene}, \ref{yalecarbonyl} and \ref{yaleazabenzene} where the presented approximate EOM methods 
obtain relative state orderings and energies to within $\sim 0.04$ eV of canonical EOM-CCSD.  


\section{Acknowledgments}

The authors acknowledge support from the U.S. Air Force Office of Scientific Research
grant FA 9550-11-1-0065 and the U.S. Army Research Office DURIP grant W911-12-1-0365 which funded access to the
University of Florida Research Computing HiPerGator high performance cluster.


\begin{thebibliography}{56}%
\makeatletter
\providecommand \@ifxundefined [1]{%
 \@ifx{#1\undefined}
}%
\providecommand \@ifnum [1]{%
 \ifnum #1\expandafter \@firstoftwo
 \else \expandafter \@secondoftwo
 \fi
}%
\providecommand \@ifx [1]{%
 \ifx #1\expandafter \@firstoftwo
 \else \expandafter \@secondoftwo
 \fi
}%
\providecommand \natexlab [1]{#1}%
\providecommand \enquote  [1]{``#1''}%
\providecommand \bibnamefont  [1]{#1}%
\providecommand \bibfnamefont [1]{#1}%
\providecommand \citenamefont [1]{#1}%
\providecommand \href@noop [0]{\@secondoftwo}%
\providecommand \href [0]{\begingroup \@sanitize@url \@href}%
\providecommand \@href[1]{\@@startlink{#1}\@@href}%
\providecommand \@@href[1]{\endgroup#1\@@endlink}%
\providecommand \@sanitize@url [0]{\catcode `\\12\catcode `\$12\catcode
  `\&12\catcode `\#12\catcode `\^12\catcode `\_12\catcode `\%12\relax}%
\providecommand \@@startlink[1]{}%
\providecommand \@@endlink[0]{}%
\providecommand \url  [0]{\begingroup\@sanitize@url \@url }%
\providecommand \@url [1]{\endgroup\@href {#1}{\urlprefix }}%
\providecommand \urlprefix  [0]{URL }%
\providecommand \Eprint [0]{\href }%
\providecommand \doibase [0]{http://dx.doi.org/}%
\providecommand \selectlanguage [0]{\@gobble}%
\providecommand \bibinfo  [0]{\@secondoftwo}%
\providecommand \bibfield  [0]{\@secondoftwo}%
\providecommand \translation [1]{[#1]}%
\providecommand \BibitemOpen [0]{}%
\providecommand \bibitemStop [0]{}%
\providecommand \bibitemNoStop [0]{.\EOS\space}%
\providecommand \EOS [0]{\spacefactor3000\relax}%
\providecommand \BibitemShut  [1]{\csname bibitem#1\endcsname}%
\let\auto@bib@innerbib\@empty
\bibitem [{\citenamefont {Dreuw}\ and\ \citenamefont
  {Head-Gordon}(2005)}]{dreuw2005}%
  \BibitemOpen
  \bibfield  {author} {\bibinfo {author} {\bibfnamefont {A.}~\bibnamefont
  {Dreuw}}\ and\ \bibinfo {author} {\bibfnamefont {M.}~\bibnamefont
  {Head-Gordon}},\ }\href@noop {} {\bibfield  {journal} {\bibinfo  {journal}
  {Chem. Rev.}\ }\textbf {\bibinfo {volume} {105}},\ \bibinfo {pages} {4009}
  (\bibinfo {year} {2005})}\BibitemShut {NoStop}%
\bibitem [{\citenamefont {Caricato}\ \emph {et~al.}(2010)\citenamefont
  {Caricato}, \citenamefont {Trucks}, \citenamefont {Frisch},\ and\
  \citenamefont {Wiberg}}]{caricato2010}%
  \BibitemOpen
  \bibfield  {author} {\bibinfo {author} {\bibfnamefont {M.}~\bibnamefont
  {Caricato}}, \bibinfo {author} {\bibfnamefont {G.~W.}\ \bibnamefont
  {Trucks}}, \bibinfo {author} {\bibfnamefont {M.~J.}\ \bibnamefont {Frisch}},
  \ and\ \bibinfo {author} {\bibfnamefont {K.~B.}\ \bibnamefont {Wiberg}},\
  }\href@noop {} {\bibfield  {journal} {\bibinfo  {journal} {J. Chem. Theory
  Comput.}\ }\textbf {\bibinfo {volume} {6}},\ \bibinfo {pages} {370} (\bibinfo
  {year} {2010})}\BibitemShut {NoStop}%
\bibitem [{\citenamefont {Szalay}\ \emph {et~al.}(2012)\citenamefont {Szalay},
  \citenamefont {Müller}, \citenamefont {Gidofalvi},\ and\ \citenamefont
  {Lischka}}]{szalay2012}%
  \BibitemOpen
  \bibfield  {author} {\bibinfo {author} {\bibfnamefont {P.~G.}\ \bibnamefont
  {Szalay}}, \bibinfo {author} {\bibfnamefont {T.}~\bibnamefont {Müller}},
  \bibinfo {author} {\bibfnamefont {G.}~\bibnamefont {Gidofalvi}}, \ and\
  \bibinfo {author} {\bibfnamefont {H.}~\bibnamefont {Lischka}},\ }\href@noop
  {} {\bibfield  {journal} {\bibinfo  {journal} {Chem. Rev.}\ }\textbf
  {\bibinfo {volume} {112}},\ \bibinfo {pages} {108} (\bibinfo {year}
  {2012})}\BibitemShut {NoStop}%
\bibitem [{\citenamefont {Bartlett}\ and\ \citenamefont
  {Musia\l}(2007)}]{bartlett2007}%
  \BibitemOpen
  \bibfield  {author} {\bibinfo {author} {\bibfnamefont {R.}~\bibnamefont
  {Bartlett}}\ and\ \bibinfo {author} {\bibfnamefont {M.}~\bibnamefont
  {Musia\l}},\ }\href@noop {} {\bibfield  {journal} {\bibinfo  {journal} {Rev.
  Mod. Phys.}\ }\textbf {\bibinfo {volume} {79}},\ \bibinfo {pages} {291}
  (\bibinfo {year} {2007})}\BibitemShut {NoStop}%
\bibitem [{\citenamefont {Sneskov}\ and\ \citenamefont
  {Christiansen}(2011)}]{sneskov2012}%
  \BibitemOpen
  \bibfield  {author} {\bibinfo {author} {\bibfnamefont {K.}~\bibnamefont
  {Sneskov}}\ and\ \bibinfo {author} {\bibfnamefont {O.}~\bibnamefont
  {Christiansen}},\ }\href@noop {} {\bibfield  {journal} {\bibinfo  {journal}
  {WIREs: Comput. Mol. Sci.}\ }\textbf {\bibinfo {volume} {2}},\ \bibinfo
  {pages} {566} (\bibinfo {year} {2011})}\BibitemShut {NoStop}%
\bibitem [{\citenamefont {Stanton}\ and\ \citenamefont
  {Bartlett}(1993)}]{stanton1993}%
  \BibitemOpen
  \bibfield  {author} {\bibinfo {author} {\bibfnamefont {J.~F.}\ \bibnamefont
  {Stanton}}\ and\ \bibinfo {author} {\bibfnamefont {R.~J.}\ \bibnamefont
  {Bartlett}},\ }\href@noop {} {\bibfield  {journal} {\bibinfo  {journal} {J.
  Chem. Phys.}\ }\textbf {\bibinfo {volume} {98}},\ \bibinfo {pages} {7029}
  (\bibinfo {year} {1993})}\BibitemShut {NoStop}%
\bibitem [{\citenamefont {{Watson Jr.}}\ \emph {et~al.}(2013)\citenamefont
  {{Watson Jr.}}, \citenamefont {Lotrich}, \citenamefont {Szalay},
  \citenamefont {Perera},\ and\ \citenamefont {Bartlett}}]{watson2013}%
  \BibitemOpen
  \bibfield  {author} {\bibinfo {author} {\bibfnamefont {T.~J.}\ \bibnamefont
  {{Watson Jr.}}}, \bibinfo {author} {\bibfnamefont {V.~F.}\ \bibnamefont
  {Lotrich}}, \bibinfo {author} {\bibfnamefont {P.~G.}\ \bibnamefont {Szalay}},
  \bibinfo {author} {\bibfnamefont {A.}~\bibnamefont {Perera}}, \ and\ \bibinfo
  {author} {\bibfnamefont {R.~J.}\ \bibnamefont {Bartlett}},\ }\href@noop {}
  {\bibfield  {journal} {\bibinfo  {journal} {J. Phys. Chem. A}\ }\textbf
  {\bibinfo {volume} {117}},\ \bibinfo {pages} {2569} (\bibinfo {year}
  {2013})}\BibitemShut {NoStop}%
\bibitem [{\citenamefont {Watts}\ and\ \citenamefont
  {Bartlett}(1994)}]{watts1994}%
  \BibitemOpen
  \bibfield  {author} {\bibinfo {author} {\bibfnamefont {J.~D.}\ \bibnamefont
  {Watts}}\ and\ \bibinfo {author} {\bibfnamefont {R.~J.}\ \bibnamefont
  {Bartlett}},\ }\href@noop {} {\bibfield  {journal} {\bibinfo  {journal} {J.
  Chem. Phys.}\ }\textbf {\bibinfo {volume} {101}},\ \bibinfo {pages} {3073}
  (\bibinfo {year} {1994})}\BibitemShut {NoStop}%
\bibitem [{\citenamefont {{Del Bene}}, \citenamefont {Watts},\ and\
  \citenamefont {Bartlett}(1997)}]{bene1997}%
  \BibitemOpen
  \bibfield  {author} {\bibinfo {author} {\bibfnamefont {J.~E.}\ \bibnamefont
  {{Del Bene}}}, \bibinfo {author} {\bibfnamefont {J.~D.}\ \bibnamefont
  {Watts}}, \ and\ \bibinfo {author} {\bibfnamefont {R.~J.}\ \bibnamefont
  {Bartlett}},\ }\href@noop {} {\bibfield  {journal} {\bibinfo  {journal} {J.
  Chem. Phys.}\ }\textbf {\bibinfo {volume} {106}},\ \bibinfo {pages} {6051}
  (\bibinfo {year} {1997})}\BibitemShut {NoStop}%
\bibitem [{\citenamefont {Watts}\ and\ \citenamefont
  {Bartlett}(1999)}]{watts1999}%
  \BibitemOpen
  \bibfield  {author} {\bibinfo {author} {\bibfnamefont {J.~D.}\ \bibnamefont
  {Watts}}\ and\ \bibinfo {author} {\bibfnamefont {R.~J.}\ \bibnamefont
  {Bartlett}},\ }\href@noop {} {\bibfield  {journal} {\bibinfo  {journal}
  {Spectrochim. Acta, Part A}\ }\textbf {\bibinfo {volume} {55}},\ \bibinfo
  {pages} {495} (\bibinfo {year} {1999})}\BibitemShut {NoStop}%
\bibitem [{\citenamefont {Paldus}\ \emph {et~al.}(1978)\citenamefont {Paldus},
  \citenamefont {\v{C}\'{i}\v{z}ek}, \citenamefont {Saute},\ and\ \citenamefont
  {Laforgue}}]{paldus1978}%
  \BibitemOpen
  \bibfield  {author} {\bibinfo {author} {\bibfnamefont {J.}~\bibnamefont
  {Paldus}}, \bibinfo {author} {\bibfnamefont {J.}~\bibnamefont
  {\v{C}\'{i}\v{z}ek}}, \bibinfo {author} {\bibfnamefont {M.}~\bibnamefont
  {Saute}}, \ and\ \bibinfo {author} {\bibfnamefont {A.}~\bibnamefont
  {Laforgue}},\ }\href@noop {} {\bibfield  {journal} {\bibinfo  {journal}
  {Phys. Rev. A}\ }\textbf {\bibinfo {volume} {17}},\ \bibinfo {pages} {805}
  (\bibinfo {year} {1978})}\BibitemShut {NoStop}%
\bibitem [{\citenamefont {Foresman}, \citenamefont {Head-Gordon},\ and\
  \citenamefont {Pople}(1992)}]{forseman1992}%
  \BibitemOpen
  \bibfield  {author} {\bibinfo {author} {\bibfnamefont {J.}~\bibnamefont
  {Foresman}}, \bibinfo {author} {\bibfnamefont {M.}~\bibnamefont
  {Head-Gordon}}, \ and\ \bibinfo {author} {\bibfnamefont {J.}~\bibnamefont
  {Pople}},\ }\href@noop {} {\bibfield  {journal} {\bibinfo  {journal} {J.
  Phys. Chem.}\ }\textbf {\bibinfo {volume} {96}},\ \bibinfo {pages} {135}
  (\bibinfo {year} {1992})}\BibitemShut {NoStop}%
\bibitem [{\citenamefont {Head-Gordon}\ \emph {et~al.}(1994)\citenamefont
  {Head-Gordon}, \citenamefont {Rico}, \citenamefont {Oumi},\ and\
  \citenamefont {Lee}}]{headgordon1994}%
  \BibitemOpen
  \bibfield  {author} {\bibinfo {author} {\bibfnamefont {M.}~\bibnamefont
  {Head-Gordon}}, \bibinfo {author} {\bibfnamefont {R.~J.}\ \bibnamefont
  {Rico}}, \bibinfo {author} {\bibfnamefont {M.}~\bibnamefont {Oumi}}, \ and\
  \bibinfo {author} {\bibfnamefont {T.~J.}\ \bibnamefont {Lee}},\ }\href@noop
  {} {\bibfield  {journal} {\bibinfo  {journal} {Chem. Phys. Lett.}\ }\textbf
  {\bibinfo {volume} {219}},\ \bibinfo {pages} {21} (\bibinfo {year}
  {1994})}\BibitemShut {NoStop}%
\bibitem [{\citenamefont {Hirata}(2005)}]{hirata2005}%
  \BibitemOpen
  \bibfield  {author} {\bibinfo {author} {\bibfnamefont {S.}~\bibnamefont
  {Hirata}},\ }\href@noop {} {\bibfield  {journal} {\bibinfo  {journal} {J.
  Chem. Phys.}\ }\textbf {\bibinfo {volume} {122}},\ \bibinfo {pages} {094105}
  (\bibinfo {year} {2005})}\BibitemShut {NoStop}%
\bibitem [{\citenamefont {Liu}\ \emph {et~al.}(2013)\citenamefont {Liu},
  \citenamefont {Ou}, \citenamefont {Alguire},\ and\ \citenamefont
  {Subotnik}}]{liu2013}%
  \BibitemOpen
  \bibfield  {author} {\bibinfo {author} {\bibfnamefont {X.}~\bibnamefont
  {Liu}}, \bibinfo {author} {\bibfnamefont {Q.}~\bibnamefont {Ou}}, \bibinfo
  {author} {\bibfnamefont {E.}~\bibnamefont {Alguire}}, \ and\ \bibinfo
  {author} {\bibfnamefont {J.~E.}\ \bibnamefont {Subotnik}},\ }\href@noop {}
  {\bibfield  {journal} {\bibinfo  {journal} {J Chem. Phys.}\ }\textbf
  {\bibinfo {volume} {138}},\ \bibinfo {pages} {221105} (\bibinfo {year}
  {2013})}\BibitemShut {NoStop}%
\bibitem [{\citenamefont {Liu}\ and\ \citenamefont
  {Subotnik}(2014{\natexlab{a}})}]{liu2014a}%
  \BibitemOpen
  \bibfield  {author} {\bibinfo {author} {\bibfnamefont {X.}~\bibnamefont
  {Liu}}\ and\ \bibinfo {author} {\bibfnamefont {J.~E.}\ \bibnamefont
  {Subotnik}},\ }\href@noop {} {\bibfield  {journal} {\bibinfo  {journal} {J.
  Chem. Theory Comput.}\ }\textbf {\bibinfo {volume} {10}},\ \bibinfo {pages}
  {1004} (\bibinfo {year} {2014}{\natexlab{a}})}\BibitemShut {NoStop}%
\bibitem [{\citenamefont {Liu}\ and\ \citenamefont
  {Subotnik}(2014{\natexlab{b}})}]{liu2014b}%
  \BibitemOpen
  \bibfield  {author} {\bibinfo {author} {\bibfnamefont {X.}~\bibnamefont
  {Liu}}\ and\ \bibinfo {author} {\bibfnamefont {J.~E.}\ \bibnamefont
  {Subotnik}},\ }\href@noop {} {\bibfield  {journal} {\bibinfo  {journal} {J.
  Chem. Theory Comput.}\ }\textbf {\bibinfo {volume} {10}},\ \bibinfo {pages}
  {1835} (\bibinfo {year} {2014}{\natexlab{b}})}\BibitemShut {NoStop}%
\bibitem [{\citenamefont {Byrd}, \citenamefont {Lotrich},\ and\ \citenamefont
  {Bartlett}(2014)}]{byrd2014-b}%
  \BibitemOpen
  \bibfield  {author} {\bibinfo {author} {\bibfnamefont {J.~N.}\ \bibnamefont
  {Byrd}}, \bibinfo {author} {\bibfnamefont {V.~F.}\ \bibnamefont {Lotrich}}, \
  and\ \bibinfo {author} {\bibfnamefont {R.~J.}\ \bibnamefont {Bartlett}},\
  }\href@noop {} {\bibfield  {journal} {\bibinfo  {journal} {J. Chem. Phys.}\
  }\textbf {\bibinfo {volume} {140}},\ \bibinfo {pages} {234108} (\bibinfo
  {year} {2014})}\BibitemShut {NoStop}%
\bibitem [{\citenamefont {Schirmer}(1982)}]{schirmer1982}%
  \BibitemOpen
  \bibfield  {author} {\bibinfo {author} {\bibfnamefont {J.}~\bibnamefont
  {Schirmer}},\ }\href@noop {} {\bibfield  {journal} {\bibinfo  {journal}
  {Phys. Rev. A}\ }\textbf {\bibinfo {volume} {26}},\ \bibinfo {pages}
  {2395–2416} (\bibinfo {year} {1982})}\BibitemShut {NoStop}%
\bibitem [{\citenamefont {Trofimov}\ \emph {et~al.}(2006)\citenamefont
  {Trofimov}, \citenamefont {Krivdina}, \citenamefont {Weller},\ and\
  \citenamefont {Schirmer}}]{trofimov2006}%
  \BibitemOpen
  \bibfield  {author} {\bibinfo {author} {\bibfnamefont {A.~B.}\ \bibnamefont
  {Trofimov}}, \bibinfo {author} {\bibfnamefont {I.~L.}\ \bibnamefont
  {Krivdina}}, \bibinfo {author} {\bibfnamefont {J.}~\bibnamefont {Weller}}, \
  and\ \bibinfo {author} {\bibfnamefont {J.}~\bibnamefont {Schirmer}},\
  }\href@noop {} {\bibfield  {journal} {\bibinfo  {journal} {Chem. Phys.}\
  }\textbf {\bibinfo {volume} {329}},\ \bibinfo {pages} {1} (\bibinfo {year}
  {2006})}\BibitemShut {NoStop}%
\bibitem [{\citenamefont {Monkhorst}(1977)}]{monkhorst1977}%
  \BibitemOpen
  \bibfield  {author} {\bibinfo {author} {\bibfnamefont {H.~J.}\ \bibnamefont
  {Monkhorst}},\ }\href@noop {} {\bibfield  {journal} {\bibinfo  {journal}
  {Int. J. Quant. Chem. Symp.}\ }\textbf {\bibinfo {volume} {11}},\ \bibinfo
  {pages} {421} (\bibinfo {year} {1977})}\BibitemShut {NoStop}%
\bibitem [{\citenamefont {Sekino}\ and\ \citenamefont
  {Bartlett}(1984)}]{sekino1984}%
  \BibitemOpen
  \bibfield  {author} {\bibinfo {author} {\bibfnamefont {H.}~\bibnamefont
  {Sekino}}\ and\ \bibinfo {author} {\bibfnamefont {R.~J.}\ \bibnamefont
  {Bartlett}},\ }\href@noop {} {\bibfield  {journal} {\bibinfo  {journal} {Int.
  J. Quantum Chem.}\ }\textbf {\bibinfo {volume} {18}},\ \bibinfo {pages} {255}
  (\bibinfo {year} {1984})}\BibitemShut {NoStop}%
\bibitem [{\citenamefont {Koch}\ and\ \citenamefont
  {J\"{o}rgensen}(1990)}]{koch1990}%
  \BibitemOpen
  \bibfield  {author} {\bibinfo {author} {\bibfnamefont {H.}~\bibnamefont
  {Koch}}\ and\ \bibinfo {author} {\bibfnamefont {P.}~\bibnamefont
  {J\"{o}rgensen}},\ }\href@noop {} {\bibfield  {journal} {\bibinfo  {journal}
  {J. Chem. Phys.}\ }\textbf {\bibinfo {volume} {93}},\ \bibinfo {pages} {3333}
  (\bibinfo {year} {1990})}\BibitemShut {NoStop}%
\bibitem [{\citenamefont {Christiansen}, \citenamefont {Koch},\ and\
  \citenamefont {J{\o}rgensen}(1995)}]{christiansen1995}%
  \BibitemOpen
  \bibfield  {author} {\bibinfo {author} {\bibfnamefont {O.}~\bibnamefont
  {Christiansen}}, \bibinfo {author} {\bibfnamefont {H.}~\bibnamefont {Koch}},
  \ and\ \bibinfo {author} {\bibfnamefont {P.}~\bibnamefont {J{\o}rgensen}},\
  }\href@noop {} {\bibfield  {journal} {\bibinfo  {journal} {Chem. Phys.
  Lett.}\ }\textbf {\bibinfo {volume} {243}},\ \bibinfo {pages} {409} (\bibinfo
  {year} {1995})}\BibitemShut {NoStop}%
\bibitem [{\citenamefont {Hirata}\ \emph
  {et~al.}(2001{\natexlab{a}})\citenamefont {Hirata}, \citenamefont {Nooijen},
  \citenamefont {Grabowski},\ and\ \citenamefont {Bartlett}}]{hirata2001a}%
  \BibitemOpen
  \bibfield  {author} {\bibinfo {author} {\bibfnamefont {S.}~\bibnamefont
  {Hirata}}, \bibinfo {author} {\bibfnamefont {M.}~\bibnamefont {Nooijen}},
  \bibinfo {author} {\bibfnamefont {I.}~\bibnamefont {Grabowski}}, \ and\
  \bibinfo {author} {\bibfnamefont {R.~J.}\ \bibnamefont {Bartlett}},\
  }\href@noop {} {\bibfield  {journal} {\bibinfo  {journal} {J. Chem. Phys.}\
  }\textbf {\bibinfo {volume} {114}},\ \bibinfo {pages} {3919} (\bibinfo {year}
  {2001}{\natexlab{a}})}\BibitemShut {NoStop}%
\bibitem [{\citenamefont {Hirata}\ \emph
  {et~al.}(2001{\natexlab{b}})\citenamefont {Hirata}, \citenamefont {Nooijen},
  \citenamefont {Grabowski},\ and\ \citenamefont {Bartlett}}]{hirata2001b}%
  \BibitemOpen
  \bibfield  {author} {\bibinfo {author} {\bibfnamefont {S.}~\bibnamefont
  {Hirata}}, \bibinfo {author} {\bibfnamefont {M.}~\bibnamefont {Nooijen}},
  \bibinfo {author} {\bibfnamefont {I.}~\bibnamefont {Grabowski}}, \ and\
  \bibinfo {author} {\bibfnamefont {R.~J.}\ \bibnamefont {Bartlett}},\
  }\href@noop {} {\bibfield  {journal} {\bibinfo  {journal} {J. Chem. Phys.}\
  }\textbf {\bibinfo {volume} {115}},\ \bibinfo {pages} {3967} (\bibinfo {year}
  {2001}{\natexlab{b}})}\BibitemShut {NoStop}%
\bibitem [{\citenamefont {Stanton}(1995)}]{stanton1995}%
  \BibitemOpen
  \bibfield  {author} {\bibinfo {author} {\bibfnamefont {J.}~\bibnamefont
  {Stanton}, \bibfnamefont {J.~F.and~Gauss}},\ }\href@noop {} {\bibfield
  {journal} {\bibinfo  {journal} {J. Chem. Phys.}\ }\textbf {\bibinfo {volume}
  {103}},\ \bibinfo {pages} {1064} (\bibinfo {year} {1995})}\BibitemShut
  {NoStop}%
\bibitem [{\citenamefont {Nooijen}\ and\ \citenamefont
  {Snijders}(1995)}]{nooijen1995}%
  \BibitemOpen
  \bibfield  {author} {\bibinfo {author} {\bibfnamefont {M.}~\bibnamefont
  {Nooijen}}\ and\ \bibinfo {author} {\bibfnamefont {J.~G.}\ \bibnamefont
  {Snijders}},\ }\href@noop {} {\bibfield  {journal} {\bibinfo  {journal} {J.
  Chem. Phys.}\ }\textbf {\bibinfo {volume} {102}},\ \bibinfo {pages} {1681}
  (\bibinfo {year} {1995})}\BibitemShut {NoStop}%
\bibitem [{\citenamefont {Gwaltney}, \citenamefont {Nooijen},\ and\
  \citenamefont {Bartlett}(1996)}]{gwaltney1996}%
  \BibitemOpen
  \bibfield  {author} {\bibinfo {author} {\bibfnamefont {S.~R.}\ \bibnamefont
  {Gwaltney}}, \bibinfo {author} {\bibfnamefont {M.}~\bibnamefont {Nooijen}}, \
  and\ \bibinfo {author} {\bibfnamefont {R.~J.}\ \bibnamefont {Bartlett}},\
  }\href@noop {} {\bibfield  {journal} {\bibinfo  {journal} {Chem. Phys.
  Lett.}\ }\textbf {\bibinfo {volume} {248}},\ \bibinfo {pages} {189} (\bibinfo
  {year} {1996})}\BibitemShut {NoStop}%
\bibitem [{\citenamefont {Bartlett}\ \emph {et~al.}(2010)\citenamefont
  {Bartlett}, \citenamefont {Musial}, \citenamefont {Lotrich},\ and\
  \citenamefont {Kus}}]{bartlett2010}%
  \BibitemOpen
  \bibfield  {author} {\bibinfo {author} {\bibfnamefont {R.~J.}\ \bibnamefont
  {Bartlett}}, \bibinfo {author} {\bibfnamefont {M.}~\bibnamefont {Musial}},
  \bibinfo {author} {\bibfnamefont {V.~F.}\ \bibnamefont {Lotrich}}, \ and\
  \bibinfo {author} {\bibfnamefont {T.}~\bibnamefont {Kus}},\ }in\ \href@noop
  {} {\emph {\bibinfo {booktitle} {Recent Progress in Coupled-Cluster
  Methods}}},\ Vol.~\bibinfo {volume} {11},\ \bibinfo {editor} {edited by\
  \bibinfo {editor} {\bibfnamefont {P.}~\bibnamefont {Carsky}}, \bibinfo
  {editor} {\bibfnamefont {J.}~\bibnamefont {Paldus}}, \ and\ \bibinfo {editor}
  {\bibfnamefont {J.}~\bibnamefont {Pittner}}}\ (\bibinfo  {publisher}
  {Springer},\ \bibinfo {address} {Dordrecht},\ \bibinfo {year} {2010})\
  Chap.~\bibinfo {chapter} {1}, pp.\ \bibinfo {pages} {1--34}\BibitemShut
  {NoStop}%
\bibitem [{\citenamefont {Paldus}, \citenamefont {\v{C}\'{i}\v{z}ek},\ and\
  \citenamefont {Shavitt}(1972)}]{paldus1972}%
  \BibitemOpen
  \bibfield  {author} {\bibinfo {author} {\bibfnamefont {J.}~\bibnamefont
  {Paldus}}, \bibinfo {author} {\bibfnamefont {J.}~\bibnamefont
  {\v{C}\'{i}\v{z}ek}}, \ and\ \bibinfo {author} {\bibfnamefont
  {I.}~\bibnamefont {Shavitt}},\ }\href@noop {} {\bibfield  {journal} {\bibinfo
   {journal} {Phys. Rev. A}\ }\textbf {\bibinfo {volume} {5}},\ \bibinfo
  {pages} {50} (\bibinfo {year} {1972})}\BibitemShut {NoStop}%
\bibitem [{\citenamefont {Bartlett}\ and\ \citenamefont
  {Noga}(1988)}]{bartlett1988}%
  \BibitemOpen
  \bibfield  {author} {\bibinfo {author} {\bibfnamefont {R.~J.}\ \bibnamefont
  {Bartlett}}\ and\ \bibinfo {author} {\bibfnamefont {J.}~\bibnamefont
  {Noga}},\ }\href@noop {} {\bibfield  {journal} {\bibinfo  {journal} {Chem.
  Phys. Lett.}\ }\textbf {\bibinfo {volume} {150}},\ \bibinfo {pages} {29}
  (\bibinfo {year} {1988})}\BibitemShut {NoStop}%
\bibitem [{\citenamefont {Bartlett}, \citenamefont {Kucharski},\ and\
  \citenamefont {J.Noga}(1989)}]{bartlett1989}%
  \BibitemOpen
  \bibfield  {author} {\bibinfo {author} {\bibfnamefont {R.~J.}\ \bibnamefont
  {Bartlett}}, \bibinfo {author} {\bibfnamefont {S.~A.}\ \bibnamefont
  {Kucharski}}, \ and\ \bibinfo {author} {\bibnamefont {J.Noga}},\ }\href@noop
  {} {\bibfield  {journal} {\bibinfo  {journal} {Chem. Phys. Lett.}\ }\textbf
  {\bibinfo {volume} {155}},\ \bibinfo {pages} {133} (\bibinfo {year}
  {1989})}\BibitemShut {NoStop}%
\bibitem [{\citenamefont {Byrd}\ \emph {et~al.}(2015)\citenamefont {Byrd},
  \citenamefont {Jindal}, \citenamefont {{Molt, Jr.}}, \citenamefont
  {Bartlett}, \citenamefont {Sanders},\ and\ \citenamefont
  {Lotrich}}]{byrd2015-a}%
  \BibitemOpen
  \bibfield  {author} {\bibinfo {author} {\bibfnamefont {J.~N.}\ \bibnamefont
  {Byrd}}, \bibinfo {author} {\bibfnamefont {N.}~\bibnamefont {Jindal}},
  \bibinfo {author} {\bibfnamefont {R.~W.}\ \bibnamefont {{Molt, Jr.}}},
  \bibinfo {author} {\bibfnamefont {R.~J.}\ \bibnamefont {Bartlett}}, \bibinfo
  {author} {\bibfnamefont {B.~A.}\ \bibnamefont {Sanders}}, \ and\ \bibinfo
  {author} {\bibfnamefont {V.~F.}\ \bibnamefont {Lotrich}},\ }\href@noop {}
  {\bibfield  {journal} {\bibinfo  {journal} {Mol. Phys.}\ }\textbf {\bibinfo
  {volume} {113}},\ \bibinfo {pages} {1} (\bibinfo {year} {2015})}\BibitemShut
  {NoStop}%
\bibitem [{\citenamefont {Shavitt}\ and\ \citenamefont
  {Bartlett}(2009)}]{shavitt2009}%
  \BibitemOpen
  \bibfield  {author} {\bibinfo {author} {\bibfnamefont {I.}~\bibnamefont
  {Shavitt}}\ and\ \bibinfo {author} {\bibfnamefont {R.~J.}\ \bibnamefont
  {Bartlett}},\ }\href@noop {} {\emph {\bibinfo {title} {Many-Body Methods in
  Chemistry and Physics}}}\ (\bibinfo  {publisher} {Cambridge},\ \bibinfo
  {address} {New York},\ \bibinfo {year} {2009})\BibitemShut {NoStop}%
\bibitem [{\citenamefont {Taube}\ and\ \citenamefont
  {Bartlett}(2009)}]{taube2009}%
  \BibitemOpen
  \bibfield  {author} {\bibinfo {author} {\bibfnamefont {A.~G.}\ \bibnamefont
  {Taube}}\ and\ \bibinfo {author} {\bibfnamefont {R.~J.}\ \bibnamefont
  {Bartlett}},\ }\href@noop {} {\bibfield  {journal} {\bibinfo  {journal} {J.
  Chem. Phys.}\ }\textbf {\bibinfo {volume} {130}},\ \bibinfo {pages} {144112}
  (\bibinfo {year} {2009})}\BibitemShut {NoStop}%
\bibitem [{\citenamefont {Schreiber}\ \emph {et~al.}(2008)\citenamefont
  {Schreiber}, \citenamefont {Silva-Junior}, \citenamefont {Sauer},\ and\
  \citenamefont {Thiel}}]{schreiber2008}%
  \BibitemOpen
  \bibfield  {author} {\bibinfo {author} {\bibfnamefont {M.}~\bibnamefont
  {Schreiber}}, \bibinfo {author} {\bibfnamefont {M.~R.}\ \bibnamefont
  {Silva-Junior}}, \bibinfo {author} {\bibfnamefont {S.~P.~A.}\ \bibnamefont
  {Sauer}}, \ and\ \bibinfo {author} {\bibfnamefont {W.}~\bibnamefont
  {Thiel}},\ }\href@noop {} {\bibfield  {journal} {\bibinfo  {journal} {J Chem.
  Phys.}\ }\textbf {\bibinfo {volume} {128}},\ \bibinfo {pages} {134110}
  (\bibinfo {year} {2008})}\BibitemShut {NoStop}%
\bibitem [{\citenamefont {Sauer}\ \emph {et~al.}(2009)\citenamefont {Sauer},
  \citenamefont {Schreiber}, \citenamefont {Silva-Junior},\ and\ \citenamefont
  {Thiel}}]{sauer2009}%
  \BibitemOpen
  \bibfield  {author} {\bibinfo {author} {\bibfnamefont {S.~P.~A.}\
  \bibnamefont {Sauer}}, \bibinfo {author} {\bibfnamefont {M.}~\bibnamefont
  {Schreiber}}, \bibinfo {author} {\bibfnamefont {M.~R.}\ \bibnamefont
  {Silva-Junior}}, \ and\ \bibinfo {author} {\bibfnamefont {W.}~\bibnamefont
  {Thiel}},\ }\href@noop {} {\bibfield  {journal} {\bibinfo  {journal} {J.
  Chem. Theory Comput.}\ }\textbf {\bibinfo {volume} {5}},\ \bibinfo {pages}
  {555} (\bibinfo {year} {2009})}\BibitemShut {NoStop}%
\bibitem [{\citenamefont {Silva-Junior}\ \emph {et~al.}(2010)\citenamefont
  {Silva-Junior}, \citenamefont {Schreiber}, \citenamefont {Sauer},\ and\
  \citenamefont {Thiel}}]{silva-junior2010}%
  \BibitemOpen
  \bibfield  {author} {\bibinfo {author} {\bibfnamefont {M.~R.~M.}\
  \bibnamefont {Silva-Junior}}, \bibinfo {author} {\bibfnamefont {M.~M.}\
  \bibnamefont {Schreiber}}, \bibinfo {author} {\bibfnamefont {S.~P. A.~S.}\
  \bibnamefont {Sauer}}, \ and\ \bibinfo {author} {\bibfnamefont {W.~W.}\
  \bibnamefont {Thiel}},\ }\href@noop {} {\bibfield  {journal} {\bibinfo
  {journal} {J Chem. Phys.}\ }\textbf {\bibinfo {volume} {133}},\ \bibinfo
  {pages} {174318} (\bibinfo {year} {2010})}\BibitemShut {NoStop}%
\bibitem [{\citenamefont {Caricato}\ \emph {et~al.}(2011)\citenamefont
  {Caricato}, \citenamefont {Trucks}, \citenamefont {Frisch},\ and\
  \citenamefont {Wiberg}}]{caricato2011}%
  \BibitemOpen
  \bibfield  {author} {\bibinfo {author} {\bibfnamefont {M.}~\bibnamefont
  {Caricato}}, \bibinfo {author} {\bibfnamefont {G.~W.}\ \bibnamefont
  {Trucks}}, \bibinfo {author} {\bibfnamefont {M.~J.}\ \bibnamefont {Frisch}},
  \ and\ \bibinfo {author} {\bibfnamefont {K.~B.}\ \bibnamefont {Wiberg}},\
  }\href@noop {} {\bibfield  {journal} {\bibinfo  {journal} {J. Chem. Theory
  Comput.}\ }\textbf {\bibinfo {volume} {7}},\ \bibinfo {pages} {456} (\bibinfo
  {year} {2011})}\BibitemShut {NoStop}%
\bibitem [{Note1()}]{Note1}%
  \BibitemOpen
  \bibinfo {note} {Partial inclusion of triples\cite {watts1996} which has all
  possible $\protect \mathaccentV {hat}05E{T}_1$ and $\protect \mathaccentV
  {hat}05E{T}_2$ to $\protect \mathaccentV {hat}05E{T}_3$ with $O(n^7)$
  scaling, but neglects the $\protect \mathaccentV {hat}05E{T}_3\rightarrow
  \protect \mathaccentV {hat}05E{T}_3$ contributions that would introduce an
  $O(n^8)$ step.}\BibitemShut {Stop}%
\bibitem [{\citenamefont {Sch\"{a}fer}, \citenamefont {Horn},\ and\
  \citenamefont {Ahlrichs}(1992)}]{schafer1992}%
  \BibitemOpen
  \bibfield  {author} {\bibinfo {author} {\bibfnamefont {A.}~\bibnamefont
  {Sch\"{a}fer}}, \bibinfo {author} {\bibfnamefont {H.}~\bibnamefont {Horn}}, \
  and\ \bibinfo {author} {\bibfnamefont {R.}~\bibnamefont {Ahlrichs}},\
  }\href@noop {} {\bibfield  {journal} {\bibinfo  {journal} {J. Chem. Phys.}\
  }\textbf {\bibinfo {volume} {97}},\ \bibinfo {pages} {2571} (\bibinfo {year}
  {1992})}\BibitemShut {NoStop}%
\bibitem [{\citenamefont {Gwaltney}\ and\ \citenamefont
  {Bartlett}(1995)}]{gwaltney1995}%
  \BibitemOpen
  \bibfield  {author} {\bibinfo {author} {\bibfnamefont {S.~R.}\ \bibnamefont
  {Gwaltney}}\ and\ \bibinfo {author} {\bibfnamefont {R.~J.}\ \bibnamefont
  {Bartlett}},\ }\href@noop {} {\bibfield  {journal} {\bibinfo  {journal}
  {Chem. Phys. Lett.}\ }\textbf {\bibinfo {volume} {241}},\ \bibinfo {pages}
  {26} (\bibinfo {year} {1995})}\BibitemShut {NoStop}%
\bibitem [{\citenamefont {Wiberg}, \citenamefont {de~Oliveira},\ and\
  \citenamefont {Trucks}(2002)}]{wiberg2002}%
  \BibitemOpen
  \bibfield  {author} {\bibinfo {author} {\bibfnamefont {K.~B.}\ \bibnamefont
  {Wiberg}}, \bibinfo {author} {\bibfnamefont {A.~E.}\ \bibnamefont
  {de~Oliveira}}, \ and\ \bibinfo {author} {\bibfnamefont {G.}~\bibnamefont
  {Trucks}},\ }\href@noop {} {\bibfield  {journal} {\bibinfo  {journal} {J.
  Phys. Chem. A}\ }\textbf {\bibinfo {volume} {106}},\ \bibinfo {pages} {4192}
  (\bibinfo {year} {2002})}\BibitemShut {NoStop}%
\bibitem [{\citenamefont {{Dunning Jr}}(1989)}]{dunning1989}%
  \BibitemOpen
  \bibfield  {author} {\bibinfo {author} {\bibfnamefont {T.}~\bibnamefont
  {{Dunning Jr}}},\ }\href@noop {} {\bibfield  {journal} {\bibinfo  {journal}
  {J. Chem. Phys.}\ }\textbf {\bibinfo {volume} {90}},\ \bibinfo {pages} {1007}
  (\bibinfo {year} {1989})}\BibitemShut {NoStop}%
\bibitem [{\citenamefont {Kendall}, \citenamefont {{Dunning Jr}},\ and\
  \citenamefont {Harrison}(1992)}]{kendall1992}%
  \BibitemOpen
  \bibfield  {author} {\bibinfo {author} {\bibfnamefont {R.}~\bibnamefont
  {Kendall}}, \bibinfo {author} {\bibfnamefont {T.}~\bibnamefont {{Dunning
  Jr}}}, \ and\ \bibinfo {author} {\bibfnamefont {R.}~\bibnamefont
  {Harrison}},\ }\href@noop {} {\bibfield  {journal} {\bibinfo  {journal} {J.
  Chem. Phys.}\ }\textbf {\bibinfo {volume} {96}},\ \bibinfo {pages} {6796}
  (\bibinfo {year} {1992})}\BibitemShut {NoStop}%
\bibitem [{\citenamefont {Goings}\ \emph {et~al.}(2014)\citenamefont {Goings},
  \citenamefont {Caricato}, \citenamefont {Frisch},\ and\ \citenamefont
  {Li}}]{goings2014}%
  \BibitemOpen
  \bibfield  {author} {\bibinfo {author} {\bibfnamefont {J.~J.}\ \bibnamefont
  {Goings}}, \bibinfo {author} {\bibfnamefont {M.}~\bibnamefont {Caricato}},
  \bibinfo {author} {\bibfnamefont {M.~J.}\ \bibnamefont {Frisch}}, \ and\
  \bibinfo {author} {\bibfnamefont {X.}~\bibnamefont {Li}},\ }\href@noop {}
  {\bibfield  {journal} {\bibinfo  {journal} {J. Chem. Phys.}\ }\textbf
  {\bibinfo {volume} {141}},\ \bibinfo {pages} {164116–10} (\bibinfo {year}
  {2014})}\BibitemShut {NoStop}%
\bibitem [{Note2()}]{Note2}%
  \BibitemOpen
  \bibinfo {note} {Where the d-aug-cc-pVDZ basis is clearly not as accurate as
  the 6-311(3+,3+)G** basis is in the case of the highest considered excited
  state of ethylene and formaldehyde.}\BibitemShut {Stop}%
\bibitem [{\citenamefont {Stanton}\ \emph {et~al.}()\citenamefont {Stanton},
  \citenamefont {Gauss}, \citenamefont {Perera}, \citenamefont {Yau},
  \citenamefont {Watts}, \citenamefont {Nooijen}, \citenamefont {Oliphant},
  \citenamefont {Szalay}, \citenamefont {Lauderdale}, \citenamefont {Gwaltney},
  \citenamefont {Beck}, \citenamefont {Balkov\'{a}}, \citenamefont {Bernholdt},
  \citenamefont {Baeck}, \citenamefont {Rozyczko}, \citenamefont {Sekino},
  \citenamefont {Huber}, \citenamefont {Pittner},\ and\ \citenamefont
  {Bartlett}}]{acesii}%
  \BibitemOpen
  \bibfield  {author} {\bibinfo {author} {\bibfnamefont {J.~F.}\ \bibnamefont
  {Stanton}}, \bibinfo {author} {\bibfnamefont {J.}~\bibnamefont {Gauss}},
  \bibinfo {author} {\bibfnamefont {S.~A.}\ \bibnamefont {Perera}}, \bibinfo
  {author} {\bibfnamefont {A.}~\bibnamefont {Yau}}, \bibinfo {author}
  {\bibfnamefont {J.~D.}\ \bibnamefont {Watts}}, \bibinfo {author}
  {\bibfnamefont {M.}~\bibnamefont {Nooijen}}, \bibinfo {author} {\bibfnamefont
  {N.}~\bibnamefont {Oliphant}}, \bibinfo {author} {\bibfnamefont {P.~G.}\
  \bibnamefont {Szalay}}, \bibinfo {author} {\bibfnamefont {W.~J.}\
  \bibnamefont {Lauderdale}}, \bibinfo {author} {\bibfnamefont {S.~R.}\
  \bibnamefont {Gwaltney}}, \bibinfo {author} {\bibfnamefont {S.}~\bibnamefont
  {Beck}}, \bibinfo {author} {\bibfnamefont {A.}~\bibnamefont {Balkov\'{a}}},
  \bibinfo {author} {\bibfnamefont {D.~E.}\ \bibnamefont {Bernholdt}}, \bibinfo
  {author} {\bibfnamefont {K.-K.}\ \bibnamefont {Baeck}}, \bibinfo {author}
  {\bibfnamefont {P.}~\bibnamefont {Rozyczko}}, \bibinfo {author}
  {\bibfnamefont {H.}~\bibnamefont {Sekino}}, \bibinfo {author} {\bibfnamefont
  {C.}~\bibnamefont {Huber}}, \bibinfo {author} {\bibfnamefont
  {J.}~\bibnamefont {Pittner}}, \ and\ \bibinfo {author} {\bibfnamefont
  {R.~J.}\ \bibnamefont {Bartlett}},\ }\href@noop {} {\enquote {\bibinfo
  {title} {{ACESII} is a product of the quantum theory project, university of
  florida},}\ }\bibinfo {note} {Integral packages included are VMOL
  (Alm\"{o}lf, J and Taylor, P. R.) VPROPS (Taylor, P. R.) and ABACUS
  (Helgaker, T. and Jensen, H. J. Aa. and J{\o}rgensen P. and Olsen J. and
  Taylor P. R.)}\BibitemShut {NoStop}%
\bibitem [{\citenamefont {Sanders}\ \emph {et~al.}()\citenamefont {Sanders},
  \citenamefont {Jindal}, \citenamefont {Byrd}, \citenamefont {Lotrich},
  \citenamefont {Lyakh}, \citenamefont {Flocke}, \citenamefont {Perera},\ and\
  \citenamefont {Bartlett}}]{aces4-electronic}%
  \BibitemOpen
  \bibfield  {author} {\bibinfo {author} {\bibfnamefont {B.~A.}\ \bibnamefont
  {Sanders}}, \bibinfo {author} {\bibfnamefont {N.}~\bibnamefont {Jindal}},
  \bibinfo {author} {\bibfnamefont {J.~N.}\ \bibnamefont {Byrd}}, \bibinfo
  {author} {\bibfnamefont {V.~F.}\ \bibnamefont {Lotrich}}, \bibinfo {author}
  {\bibfnamefont {D.}~\bibnamefont {Lyakh}}, \bibinfo {author} {\bibfnamefont
  {N.}~\bibnamefont {Flocke}}, \bibinfo {author} {\bibfnamefont
  {A.}~\bibnamefont {Perera}}, \ and\ \bibinfo {author} {\bibfnamefont {R.~J.}\
  \bibnamefont {Bartlett}},\ }\href@noop {} {\enquote {\bibinfo {title} {Aces4
  pre-alpha release},}\ }\bibinfo {note}
  {{https://github.com/UFParLab}}\BibitemShut {NoStop}%
\bibitem [{\citenamefont {L\"{o}wdin}(1963)}]{lowdin1963}%
  \BibitemOpen
  \bibfield  {author} {\bibinfo {author} {\bibfnamefont {P.-O.}\ \bibnamefont
  {L\"{o}wdin}},\ }\href@noop {} {\bibfield  {journal} {\bibinfo  {journal} {J.
  Mol. Spectrosc.}\ }\textbf {\bibinfo {volume} {10}},\ \bibinfo {pages} {12}
  (\bibinfo {year} {1963})}\BibitemShut {NoStop}%
\bibitem [{\citenamefont {Nooijen}\ and\ \citenamefont
  {Bartlett}(1997{\natexlab{a}})}]{nooijen1997a}%
  \BibitemOpen
  \bibfield  {author} {\bibinfo {author} {\bibfnamefont {M.}~\bibnamefont
  {Nooijen}}\ and\ \bibinfo {author} {\bibfnamefont {R.~J.}\ \bibnamefont
  {Bartlett}},\ }\href@noop {} {\bibfield  {journal} {\bibinfo  {journal} {J.
  Chem. Phys.}\ }\textbf {\bibinfo {volume} {106}},\ \bibinfo {pages} {6441}
  (\bibinfo {year} {1997}{\natexlab{a}})}\BibitemShut {NoStop}%
\bibitem [{\citenamefont {Nooijen}\ and\ \citenamefont
  {Bartlett}(1997{\natexlab{b}})}]{nooijen1997b}%
  \BibitemOpen
  \bibfield  {author} {\bibinfo {author} {\bibfnamefont {M.}~\bibnamefont
  {Nooijen}}\ and\ \bibinfo {author} {\bibfnamefont {R.~J.}\ \bibnamefont
  {Bartlett}},\ }\href@noop {} {\bibfield  {journal} {\bibinfo  {journal} {J.
  Chem. Phys.}\ }\textbf {\bibinfo {volume} {106}},\ \bibinfo {pages} {6449}
  (\bibinfo {year} {1997}{\natexlab{b}})}\BibitemShut {NoStop}%
\bibitem [{\citenamefont {Nooijen}\ and\ \citenamefont
  {Bartlett}(1997{\natexlab{c}})}]{nooijen1997c}%
  \BibitemOpen
  \bibfield  {author} {\bibinfo {author} {\bibfnamefont {M.}~\bibnamefont
  {Nooijen}}\ and\ \bibinfo {author} {\bibfnamefont {R.~J.}\ \bibnamefont
  {Bartlett}},\ }\href@noop {} {\bibfield  {journal} {\bibinfo  {journal} {J.
  Chem. Phys.}\ }\textbf {\bibinfo {volume} {107}},\ \bibinfo {pages} {6812}
  (\bibinfo {year} {1997}{\natexlab{c}})}\BibitemShut {NoStop}%
\bibitem [{\citenamefont {Sous}, \citenamefont {Goel},\ and\ \citenamefont
  {Nooijen}(2014)}]{sous2014}%
  \BibitemOpen
  \bibfield  {author} {\bibinfo {author} {\bibfnamefont {J.}~\bibnamefont
  {Sous}}, \bibinfo {author} {\bibfnamefont {P.}~\bibnamefont {Goel}}, \ and\
  \bibinfo {author} {\bibfnamefont {M.}~\bibnamefont {Nooijen}},\ }\href@noop
  {} {\bibfield  {journal} {\bibinfo  {journal} {Mol. Phys.}\ }\textbf
  {\bibinfo {volume} {112}},\ \bibinfo {pages} {616–638} (\bibinfo {year}
  {2014})}\BibitemShut {NoStop}%
\bibitem [{\citenamefont {Watts}\ and\ \citenamefont
  {Bartlett}(1996)}]{watts1996}%
  \BibitemOpen
  \bibfield  {author} {\bibinfo {author} {\bibfnamefont {J.~D.}\ \bibnamefont
  {Watts}}\ and\ \bibinfo {author} {\bibfnamefont {R.~J.}\ \bibnamefont
  {Bartlett}},\ }\href@noop {} {\bibfield  {journal} {\bibinfo  {journal}
  {Chem. Phys. Lett.}\ }\textbf {\bibinfo {volume} {258}},\ \bibinfo {pages}
  {581–588} (\bibinfo {year} {1996})}\BibitemShut {NoStop}%
\end{thebibliography}

%

\end{document}